\documentclass[12pt]{article}
\usepackage{setspace}
\usepackage{amsmath,amssymb,mathrsfs}
\usepackage{color}
\usepackage{hyperref}
\usepackage{graphicx,epsfig}

\numberwithin{equation}{section}
\begin{document}
\setlength{\topmargin}{-1cm} 
\setlength{\oddsidemargin}{-0.25cm}
\setlength{\evensidemargin}{0cm}
\newcommand{\e}{\epsilon}
\newcommand{\beq}{\begin{equation}}
\newcommand{\eeq}[1]{\label{#1}\end{equation}}
\newcommand{\bea}{\begin{eqnarray}}
\newcommand{\eea}[1]{\label{#1}\end{eqnarray}}
\renewcommand{\Im}{{\rm Im}\,}
\renewcommand{\Re}{{\rm Re}\,}
\newcommand{\diag}{{\rm diag} \, }
\newcommand{\Tr}{{\rm Tr}\,}
\def\draftnote#1{{\color{red} #1}}
\def\bldraft#1{{\color{blue} #1}}
\def\n{n \cdot v}
\def\ni{n\cdot v_I}

\begin{titlepage}
\begin{center}

\vskip 4 cm

{\Large \bf Intertwiners for D=3 Gauge  Theories}

\vskip 1 cm

\large {\bf P.A. Grassi}$^{a,b}$~\footnote{E-mail: \href{mailto:pietro.grassi@uniupo.it}{pietro.grassi@uniupo.it}} 
and   
\large {\bf E.M.G. Land\`o}$^{c}$~\footnote{E-mail:\href{mailto:emglandro@uninsubria.it}{emglandro@uninsubria.it}}

\vskip .75 cm

{ $^{(a)}$\small \it Dipartimento di Scienze e Innovazione Tecnologica (DiSIT),\\
Universit\`a del Piemonte Orientale, Viale T. Michel, 11, 15121 Alessandria, Italy}

{$^{(b)}$\small
 \it INFN,  {\it Sezione di Torino,} 
 {\it Via P. Giuria, 1, 10125, Torino, Italy}}

{
$^{(c)}$\small \it Università dell' Insubria, Via Vallareggio, 11, 22100 Como, Italy}

\vskip 1.25 cm

\begin{abstract}
\noindent  
We apply the intertwiner operator method of arXiv:2411.08865 to topological field theories, including BF theories, Chern–Simons theory, and three-dimensional gravity. We construct the operator on foliated manifolds while preserving covariance on the Cauchy surface, and compare canonical and holomorphic quantization, providing the intertwiner in both frameworks. For three-dimensional gravity, we present both covariant and time-gauge formulations, analyze the constraints, and construct the corresponding intertwiner. As an application, we derive the path ordering of Wilson loops in Chern–Simons theory. The study of observables is left for future work.

\end{abstract}

\end{center}

\end{titlepage}
\tableofcontents

\vskip 1 cm



\section{Introduction}

In Ref. \cite{dirac} , P. A. M. Dirac develops a gauge-invariant formulation of quantum electrodynamics. The fermionic field $\psi(x)$ with electric charge $e$ is considered, and the construction of a dressing operator $C[A](x)$ at fixed time is proposed, such that
\begin{eqnarray}
    \label{DIA}
    \psi^{g.i.}(x) = e^{i e C[A](x)}\psi(x)
\end{eqnarray}
Here, $C[A](x)$ is a local functional of the gauge potential $A$ defined over the entire three-dimensional space and depends on the spatial position $x$. It can be expressed as
\begin{eqnarray}
    \label{DIB}
    C[A](x) = \int d^3y G_\mu(x-y) A^\mu(y)  
\end{eqnarray}
where the kernel $G_\mu(x-y)$ satisfies the condition $\partial^\mu G_\mu(x-y) = \delta^3(x-y)$. One possible solution to this equation is given by $G_\mu(x-y) = \frac{1}{4\pi} \frac{(x-y)_\mu }{||x-y||^3}$ which corresponds to a particular choice among several admissible solutions satisfying simple boundary conditions.

At the quantum level, gauge symmetry is replaced by BRST symmetry, which is obtained by promoting the gauge symmetry to a rigid anticommuting symmetry and by introducing unphysical ghost fields in order to preserve the unitarity of the theory once a gauge fixing is implemented and the Hilbert space is constructed. Within the canonical quantization framework, the BRST charge for QED is given by the operator  
\begin{eqnarray}
    \label{DIC}
     Q= \int d^3x\left( c\partial_i E^i -\rho \pi^b -ie c \psi^\dagger \psi \right)  .
\end{eqnarray}
where $E^i$ is the electric field conjugated to the potential $A_i$, $c$ is the ghost field, $\rho$ is the Nakanishi-Lautrup field, and $\pi^b$ is the conjugate momenta of the anti-ghost $b$. Computing the BRST variation of \eqref{DIB} yields $[Q, C[A](x)] = c(x)$ which constitutes a crucial transformation rule for the construction of a map to gauge-invariant observables.

In Ref. \cite{grassi-porrati}, a new operator $\Omega$ is introduced, which intertwines an “asymptotic” BRST operator $Q_0$ with the full BRST charge $Q$. Setting the coupling constant $e$ to zero in the BRST charge results in the elimination of the non-linear contribution, yielding the simplified operator $Q_0$. With respect to $Q_0$, the fermionic field $\psi$ and its conjugate $\psi^\dagger$ are invariant and therefore belong to the local cohomology of $Q_0$.

The map relating the full BRST charge to the simplified one, together with the corresponding map between the fermionic fields, is given by
\begin{eqnarray}
    \label{DID}
    Q = \Omega^{-1} Q_0 \Omega\,,  ~~~~~
    \psi^{g.i.} = \Omega \psi 
\end{eqnarray}
Ref. \cite{grassi-porrati} also provides a detailed discussion of the construction of this operator, leading to the result
\begin{eqnarray}
\label{newE}
\Omega(0) = exp \left( - e 
\int d^3y (\psi^\dagger \psi)(y)  \int d^3 z \frac{(y-z)_i}{||y-z||^3}A^i(z)  \right) 
\end{eqnarray}
which coincides with the Dirac operator acting on fermions. 

In \cite{grassi-porrati}, Yang–Mills theory and gravity were studied within the framework of canonical quantization, with the construction carried out for specific background geometries. Nevertheless, it is of interest to develop a more general framework that accommodates different quantization schemes. To this end, we focus on simplified models that allow us to investigate the effects of various quantization procedures. In particular, we consider topological theories with a limited number of degrees of freedom and present the technical details required to construct the intertwiner along the lines introduced in the aforementioned work.

In Sec. 2, we consider $D=3$ Maxwell theory coupled to scalar fields and perform a complete analysis of the Hamiltonian constraints. In Sec. 3, we study Chern–Simons theory with a non-Abelian gauge group, analysing the model both in canonical quantization and in holomorphic quantization. The difference between these two approaches is relevant for the construction of the intertwiner $\Omega$ in more general contexts. In the holomorphic framework, the construction is rather straightforward, whereas in canonical quantization it becomes more involved. Indeed, according to the filtration used in the construction, the BRST charge contains both positive- and negative-charge components, and the definition of the intertwiner requires a suitable modification. In the present cases, this difficulty is overcome by exploiting a key property of topological field theories, namely the vanishing of local BRST cohomologies.
In the meanwhile, we show howt the physical operators (elements of the cohomology) appear in an other sector of the functional space. 

Following the Chern–Simons analysis, in Sec. 5, we construct the intertwiner for $D=3$ BF theories. In this case, the holomorphic quantization formalism developed for Chern–Simons theory can be directly applied. Nevertheless, one may also include an additional cosmological term in the action; according to the chosen filtration, this term carries negative charge and therefore requires special care in the construction.

Finally, in Sec. 6, we analyse $D=3$ gravity. We consider both canonical quantization and time-gauge fixing, including the case with a cosmological constant. In the absence of a cosmological constant, the construction closely parallels that of BF theories. When a cosmological constant is included, however, quantization can be performed either via holomorphic methods, as in Chern–Simons theory, or through canonical quantization, which in this case requires a two-step procedure to construct the intertwiner.

\section{The General Construction of the Intertwiner}

In this section, the construction of an \emph{intertwiner} operator between two BRST charges is summarized, as presented in~\cite{grassi-porrati}. Recently, a new application of the technique to $\bar TT$ deformations of 2D conformal field theory has been presented in \cite{deSabbata:2026aum}. There, further details of the construction have been deepened (see, for example, the first section and the appendix).

One starts from a nilpotent BRST operator $Q = Q_0 + Q_I$, where $Q_0$ is itself nilpotent, $Q_0^2 = 0$. The nilpotency of $Q$ then implies $[ Q_I, Q_0 ]_+ + Q_I^2 = 0$.
Another essential ingredient is the operator $R$, which satisfies
\begin{eqnarray}
    \label{opA1}
    [Q_0, R]_+ = i S\,, ~~~~~~ [S, Q_0] =0 \,,
\end{eqnarray}
where  $S$ is a 
counting operator. 
An operator intertwining the “simple” BRST charge $Q_0$ with the full BRST charge $Q$ can be constructed, provided that $Q$ admits a finite decomposition into terms which, for an appropriate choice of the sign of $R$, carry non-negative $S$-charge as
\begin{equation}
    \label{mass17}
    Q=Q_0 +\sum_{0<n<N}Q_n  \qquad [S,Q_n]=n Q_n \, , \quad N\in \mathbb{N}.
\end{equation}
In Appendix A, the theorem establishing the equivalence between the cohomologies of the BRST charges $Q_0$ and $Q$ is reviewed, based on a suitable filtration. The construction of the operator $R$ involves a certain degree of arbitrariness, as it fixes the grading associated with the filtration operator $S$. Nevertheless, as will be shown in the following sections, some convenient quantization schemes naturally lead to well-defined operators $R$ and $S$ for which the BRST charge contains components with negative $S$-charge.

 Consider now the case in which the BRST charge contains a component with negative $S$-charge, and assume that the cohomology $H(Q_{\geq 0})$ is trivial. One then has
\begin{eqnarray}
\label{negA}
Q = Q_{\geq 0} + Q_{-N}\,. 
\end{eqnarray}
The nilpotency of $Q$ implies the following relations (nilpotency conditions):
\begin{equation}
    Q_{\geq 0}^2=0 \quad [Q_{\geq 0}, Q_{-N}]_+ =0 \quad [Q_{-N},Q_{-N} ]_+=0
\end{equation}
where it is assumed that $Q_{\geq 0}^2=0$ holds independently of the presence of $Q_{-N}$.
The second condition implies that
\begin{eqnarray}
    \label{negB}
    Q_{-N} = [Q_{\geq 0}, \Sigma^{(0)}_{-N}]
\end{eqnarray}
where $ \Sigma^{(0)}_{-N}$ is an expression polynomial on the fields and their derivatives carrying at least $S$-charge $-N$, it may contain pieces with higher negative charges. One may then attempt to rewrite Eq.~\eqref{negA} by means of a similarity transformation generated by $\Sigma^{(0)}_{-N}$, namely 
\begin{eqnarray}
    \label{negC}
    Q = e^{-\Sigma^{(0)}_{-N}} Q_{\geq 0} e^{-\Sigma^{(0)}_{-N}} = 
    Q_{\geq 0} + [Q_{\geq 0}, \Sigma^{(0)}_{-N}] + \frac12 [[Q_{\geq 0}, \Sigma^{(0)}_{-N}],\Sigma^{(0)}_{-N}] + \dots  
\end{eqnarray}
The second term already reproduces the decomposition \eqref{negA}. One must therefore verify that the next contribution, $[[Q_{\geq 0}, \Sigma^{(0)}_{-N}],\Sigma^{(0)}_{-N}]$, can be consistently removed.
To this end, note that the last nilpotency condition yields, 
\begin{eqnarray}
    \label{negD}
    0= [Q_{-N}, Q_{-N}]_+ = \Big[ 
    [Q_{\geq 0}, \Sigma^{(0)}_{-N}], [Q_{\geq 0}, \Sigma^{(0)}_{-N}]
        \Big]_+ = \Big[Q_{\geq 0}, [Q_{-N},\Sigma^{(0)}_{-N}]\Big]_+ 
\end{eqnarray}
Since the cohomology of $Q_{\geq 0}$ is assumed to be trivial, this implies 
\begin{eqnarray}
    \label{negE}
    [Q_{-N},\Sigma^{(0)}_{-N}] = [Q_{\geq 0}, \Sigma_{-2N}^{(1)}].
\end{eqnarray}
where $ \Sigma_{-2N}^{(1)}$ is an expression carrying at least negative charge $-2N$, but it may contain addition higher negtive pieces. 
As a consequence,
\begin{eqnarray}
    \label{negF}
    Q = Q_{\geq 0} + 
    \Big[Q_{\geq 0}, \left( \Sigma^{(0)}_{-N} + \frac12 \Sigma_{-2N}^{(1)} \right)\Big ] + \dots
\end{eqnarray}
The additional term can therefore be reabsorbed by redefining  $\Sigma^{(0)}_{-N} \rightarrow \Sigma^{(0)}_{-N} + \frac12 \Sigma^{(1)}_{-2N}$. Iterating this procedure, higher-order contributions with charges $-3N$ (and higher negative charges) are generated. Indeed, exponentiating back the new term, one ends up with the following three (at least) $-3N$-charged terms 
\begin{eqnarray}
    \label{negG}
    \Big[\Big[\Big[Q_{\geq 0}, 
    \Sigma^{(0)}_{-N}\Big], \Sigma^{(0)}_{-N}\Big], \Sigma^{(0)}_{-N}\Big]\,, ~~~~~
    \Big[\Big[Q_{\geq 0}, \Sigma^{(0)}_{-N}\Big], \Sigma^{(1)}_{-2N}\Big]\,, ~~~~~
    \Big[\Big[Q_{\geq 0}, \Sigma^{(1)}_{-2N}\Big], \Sigma^{(0)}_{-N}\Big]\,,
\end{eqnarray}
The first one can be rewritten as follows 
\begin{eqnarray}
    \label{negH}
    \Big[\Big[\Big[Q_{\geq 0}, 
    \Sigma^{(0)}_{-N}\Big], \Sigma^{(0)}_{-N}\Big], \Sigma^{(0)}_{-N}\Big] = 
    \Big[\Big[Q_{-N}, \Sigma^{(0)}_{-N}\Big], \Sigma^{(0)}_{-N}\Big] = 
    \Big[\Big[Q_{\geq 0}, \Sigma^{(1)}_{-2N}\Big],\Sigma^{(0)}_{-N} \Big]
\end{eqnarray}
which has the form of the last one of \eqref{negG}. 
The second expression of \eqref{negG} can be rewritten as follows 
\begin{eqnarray}
    \label{negI}
    \Big[\Big[Q_{\geq 0}, \Sigma^{(0)}_{-N}\Big], \Sigma^{(1)}_{-2N}\Big] = \Big[Q_{-N}, \Sigma^{(1)}_{-2N}\Big]
\end{eqnarray}
Acting with $Q_0$, one finds
\begin{eqnarray}
    \label{negJ}
    \Big[Q_{\geq 0}, \Big[Q_{-N}, \Sigma^{(1)}_{-2N}\Big]\Big]_+ = \Big[Q_{-N}, \Big[Q_{\geq 0}, \Sigma^{(1)}_{-2N}]]_+ = \Big[Q_{-N}, Q_{-N}\Big]_+ =0
\end{eqnarray}
where the anticommutation properties of $Q_{-N}$ and $Q_{\geq 0}$ have been used. Since the cohomology of $Q_{\geq 0}$ is trivial, it follows that
\begin{eqnarray}
    \label{negHA}
    \Big[Q_{-N}, \Sigma^{(1)}_{-2N}\Big] = 
    \Big[Q_{\geq 0}, \Sigma^{(2)}_{-3N}\Big]
\end{eqnarray}
where $\Sigma^{(2)}_{-3N}$ is an expression carrying at least $-3N$ charge, but it may contain higher negative charges. 

The last expression in  \eqref{negG} can be rewritten using the Jacobi identities as follows 
\begin{eqnarray}
    \label{negK}
    \Big[\Big[Q_{\geq 0}, \Sigma^{(1)}_{-2N}\Big], \Sigma^{(0)}_{-N}\Big] = \Big[Q_{\geq 0}, \Big[\Sigma^{(2)}_{-2N}, \Sigma^{(1)}_{-N}\Big],\Big] - \Big[\Big[Q_{\geq 0}, \Sigma^{(0)}_{-N}\Big], \Sigma^{(1)}_{-2N}\Big] 
\end{eqnarray}
Once again, the total expression can be written as $Q_{\geq 0}$ of a new term with charge $-3N$ (and possibly higher negative terms). 

By the same argument, and using the triviality of the $Q_{\geq 0}$ cohomology, these terms can always be reabsorbed by suitable redefinitions of the exponent appearing in the intertwiner $\Omega$.

\section{3D Scalar - Maxwell theory}

Before moving to topological models such as Chern-Simons and 3D gravity, it is useful to review the case of 3D Maxwell theory in order to fix the notation and introduce the geometric language that will be used below. 
We consider the Maxwell action: 
\begin{eqnarray}
    \label{MAA}
    S = - \frac14\int_\mathcal{M} dA \wedge \star dA = 
    \int_\mathcal{M} \left(B \wedge dA + \frac12 B \wedge \star_3 B\right) 
\end{eqnarray}

The field $B$ is a 1-form and $\star_3$ is the Hodge dual on the 3D manifold $\mathcal{M}$. The manifold $\mathcal{M}$ is foliated into a Riemann surface $\Sigma$ and a line, so that $\mathbb{R}\times \Sigma$.

The vector potential $A$ is then decomposed in $1+2$ dimensions as $A = A_0 dt + {\tilde A}$  where $A_0$ is a zero-form and ${\tilde A}$ is a $2$-D one-form that is embedded in the Riemann surface. In the same way, we decompose the $1$-form as $B = B_0 dt + \tilde B$, where $B_0$ is a 0-form and $\tilde B$ is a 2-dimensional 1-form. Then, the Hodge dual is also decomposed as
    \begin{eqnarray}
    \label{MAB3}
    \star_3 B = \star_2 \tilde B \wedge dt + \star_2 B_0
\end{eqnarray}
This decomposition is not the most general. In curved spacetime, the situation becomes more involved, but it holds at least for $\Sigma = \mathbb{R}^2$

Inserting the decomposition in the action \eqref{MAA}, the action becomes 
\begin{equation}
\label{MABA}
    S = \int dt \int_\Sigma \Big(\tilde B\wedge(d_2A_0-\dot{{\tilde A}} ) +B_0d_2{\tilde A} +\frac{1}{2}(\tilde B\wedge\star_2\tilde B +B_0\star_2B_0)\Big) 
\end{equation}
which is gauge invariant under $\delta A_0 = \dot \lambda$ 
and $\delta {\tilde A} = d_2 \lambda$, where $d_2$ is the differential on $\Sigma$. 

The momenta are given by 
\begin{eqnarray}
    \label{MAC}
    \pi^{A_0} =0\,, ~~~~~
    \pi^{{\tilde A}} =  + \tilde B\,, ~~~~
    \pi^{B_0} =0\,, ~~~~~ 
    \pi^{\tilde B} =0\,, ~~~~~ 
\end{eqnarray}
where the only non-trivial momentum is related to the fields $\tilde A$ while the others are vanishing. The constraints $ \pi^{{\tilde A}} - \tilde B = 0$ and $ \pi^{\tilde B} =0$ are second-class constraints, since they have a non-vanishing  canonical commutator 
\begin{eqnarray}
    \label{MACA}
    [\pi^{{\tilde A}}(x) + \tilde B(x), \pi^{\tilde B}(y) ] = i\delta^2(x-y)
\end{eqnarray}

It is possible then to remove the fields $(\tilde B, \pi^{\tilde B})$ and indeed, express the Hamiltonian without those fields. 

The Hamiltonian is given by 
\begin{eqnarray}
    \label{MAD3}
    H = - \int_{\Sigma} \left( 
     \pi^{{\tilde A}} \wedge d_2 A_0 + B_0 d_2 {\tilde A} + \frac12 B_0  \star_2 B_0 + \frac12 \pi^{{\tilde A}}  \wedge \star_2 \pi^{{\tilde A}}  \right)  
\end{eqnarray}
The Hamilton-Jacobi equations can be written in the following way: 
\begin{eqnarray}
    \label{MAE3}
    \dot \pi^{A_0} &=& -\frac{\partial H}{\partial A_0} = -d_2 \pi^{{\tilde A}} =0\,, \nonumber \\
    \dot \pi^{{\tilde A}} &=& -\frac{\partial H}{\partial {\tilde A}} = -d_2 B_0 \,, \nonumber \\
    \dot \pi^{B_0} &=& -\frac{\partial H}{\partial B_0} = d_2 {\tilde A} + \star_2 B_0 = 0 \,. 
\end{eqnarray}
This set of equations constitutes the secondary constraints. The first one is a first-class constraint, since it commutes with all the others (as a consequence of the nilpotency of $d_2$). By contrast, the third equation is a {\it second-class} constraint, as it does not commute with $\pi^{B_0} =0$ given in \eqref{MAC}. This allows us to eliminate the field $B_0$; indeed, it can be integrated out, leading to the Hamiltonian
\begin{eqnarray}
    \label{MADA}
    H = - \int_{\Sigma} \left( 
     \pi^{{\tilde A}} d_2 A_0 - \frac12  d_2 {\tilde A}  \star_2d_2 {\tilde A}+ \frac12 \pi^{{\tilde A}}  \wedge \star_2 \pi^{{\tilde A}}  \right)  
\end{eqnarray}
The first one, $d_2 \pi^{{\tilde A}} =0$ is the Gauss law. The third  equation can be solved 
$B_0 = - \star_2 d_2 {\tilde A}$, and inserted it into the second equation
\begin{eqnarray}
    \label{MAEA}
    \dot \pi^{{\tilde A}} = - d_2 \star_2 d_2 {\tilde A} \,, 
\end{eqnarray}
The rest of the Hamilton-Jacobi equations are 
\begin{eqnarray}
    \label{MAF}
    \dot {{\tilde A}} = - d_2 A_0 - \star_2 \pi^{{\tilde A}} \Longrightarrow  \pi^{{\tilde A}} = - \star_2 d_2 A_0 - \star_2 \dot {{\tilde A}} 
\end{eqnarray}
which expresses the conjugate momentum in terms of the time derivative of the field $\tilde A$. The system still exhibits a gauge symmetry, which can be used to fix $A_0 =0$ and thus obtain the following equation of motion
\begin{eqnarray}
    \label{MAG}
    \partial_0^2 {\tilde A} - d_2^\dagger d_2 {\tilde A} =0\,, ~~~~  
\end{eqnarray}
with the constraint $d_2^\dagger \tilde A=0$. 
Since the constraint $d_2 \pi^{{\tilde A}} =0$ is a first-class constraint, it can be implemented at the quantum level in the BRST charge, introducing the ghost field 0-form $c$ and its 0-form Lagrange multipliers $A_0$, the  BRST doublets 2-form $\rho$ and its corresponding 2-form ghost to implement the
gauge fixing condition and the corresponding ghost dynamics. Those degrees of freedom are then quantized as follows:
\begin{align}
    &[A_0(x),\rho(y)] \;\;= i \delta^2(x-y)Vol_\Sigma\,, ~\\
    &[c(x),p(y)]_+ \;=  \delta^2(x-y)Vol_\Sigma\,, ~
    &[b(x),\pi^b(y)]_+ =  \delta^2(x-y)Vol_\Sigma
\end{align}
where $Vol_\Sigma$  is the $2$-form volume of the Riemann surface $\Sigma$. 
The BRST charge is therefore 
\begin{eqnarray}
    \label{MAGA}
    Q = \int_{\Sigma} c\,  d_2 \pi^{{\tilde A}} +\pi^b\rho
\end{eqnarray}
which generate the BRST transformations $[Q, {\tilde A}] = d_2 c$. 
One readily verifies that the BRST operator is fully linear; as a consequence, the associated intertwiner is constant in time.

\subsection{Intertwiner for 3D Scalar QED}
In order to obtain a non-trivial intertwiner, a minimal coupling between the Maxwell action and a complex scalar field is introduced. The action then takes the form 
\begin{eqnarray}
    \label{MAA-mat}
    S = 
    \int B \wedge dA + \frac12 B \wedge \star_3 B - D\phi \wedge\star_3 {D\phi}^\dagger 
\end{eqnarray}
with 
\begin{equation}
    D\phi = d\phi +ieA\phi \quad\quad {D\phi}^\dagger = d{\phi}^\dagger -ieA{\phi}^\dagger
\end{equation}
Decomposing such an action in $2+1$ formalism, we obtain 
\begin{align}
    \label{MABA-mat}
    S = \int dt \int_\Sigma& \tilde B\wedge(d_2A_0-\dot{{\tilde A}} ) +B_0d_2{\tilde A} +\frac{1}{2}(\tilde B\wedge\star_2\tilde B +B_0\star_2B_0) +\\& D_2\phi\wedge\star_2 {D_2\phi}^\dagger + (\dot\phi+ieA_0\phi)\star_2(\dot{\phi}^\dagger-ieA_0  \phi^\dagger)
\end{align}
where the operator $D$ is the covariant derivative, and it is given by
\begin{equation}
    D_2\phi = d_2\phi +ie{\tilde A}\phi \quad\quad {D_2 \phi}^\dagger = d_2 \phi^\dagger +ie{\tilde A}\overline \phi
\end{equation}
The momenta are therefore
\begin{eqnarray}
    \label{MAC-mat}
    \pi^{A_0} =0\,, ~~~
    \pi^{{\tilde A}} = \tilde B\,, ~~~
    \pi^{B_0} =0\,, ~~~ 
    \pi^{\tilde B} =0\,, ~~~ 
    \pi^{\phi} =\star_2(\dot{\phi}^\dagger-ieA_0 \phi^\dagger)\,, ~~~ 
    \pi^{\phi^\dagger} =\star_2(\dot{\phi}+ieA_0 \phi)\, ~~~ 
\end{eqnarray}
Following the same reasoning as in the free Maxwell case, the Hamiltonian takes the form
\begin{align}
    \label{MADA-mat}
    H =  \int_{\Sigma}     &\pi^{{\phi}^\dagger} \star_2\pi^\phi-\pi^{{\tilde A}} \wedge d_2 A_0 - B_0 d_2 {\tilde A} - \frac12 B_0  \star_2 B_0 - \frac12 \pi^{{\tilde A}}  \wedge \star_2 \pi^{{\tilde A}}\\&-D_2\phi\wedge\star_2{D_2\phi}^\dagger +ieA_0(\pi^{{\phi}^\dagger} {\phi}^\dagger-\pi^\phi\phi)-D_2\phi\star_2{D_2\phi}^\dagger  \nonumber 
\end{align}
with the corresponding Hamilton–Jacobi equations given by
\begin{eqnarray}
    \label{MAE3-mat}
    \dot \pi^{A_0} &=&  -(d_2 \pi^{{\tilde A}} +J^0) =0\,, \nonumber \\
    \dot \pi^{{\tilde A}} &=& - d_2 B_0 +ie(\phi\star_2d_2{\phi}^\dagger-{\phi}^\dagger\star_2d_2\phi)-(ie)^2{\tilde A} \phi {\phi}^\dagger\,, \nonumber \\
    \dot \pi^{B_0} &=&  d_2 {\tilde A} + \star_2B_0=0 \\
    \dot \pi^{\phi} &=& -D_2\wedge\star_2 {D_2\phi}^\dagger+ieA_0\pi^\phi\,,\nonumber\\
     \dot \pi^{\phi^\dagger} &=& -{D_2}^\dagger\star_2D_2\phi-ieA_0\pi^{{\phi}^\dagger} \,\nonumber
\end{eqnarray}
where we have introduced the shorthand $J^0 =ie(\pi^{{\phi}^\dagger} {\phi}^\dagger-\pi^\phi\phi)$.
The second, the fourth, and the fifth equations are actually equations of motion for the complex field and the gauge field ${\tilde A}$. The second equation is a second-class constraint that, once again, allows us to eliminate $B_0$. The first equation is a first-class constraint and hence the generator of the gauge symmetry.
The BRST charge can be written as:
\begin{eqnarray}
    \label{MAGA-mat}
    Q = \int_{\Sigma} c\,  \big(d_2 \pi^{{\tilde A}}+J^0\big) +\pi^b\rho = \int_{\Sigma} c\,  \big(d_2 \pi^{{\tilde A}}+ie(\pi^{{\phi}^\dagger}{\phi}^\dagger-\pi^\phi\phi)\big) +\pi^b\rho
\end{eqnarray}
In this case, the BRST charge acquires a non-linear contribution arising from the matter sector of the action.

The BRST charge can thus be decomposed into the following operators:
\begin{align}
    Q_0 = \int_{\Sigma} c\,  d_2 \pi^{{\tilde A}} +\pi^b\rho \quad\quad Q_1 =  ie\int_{\Sigma} c\, (\pi^{{\phi}^\dagger}{\phi}^\dagger-\pi^\phi\phi) 
\end{align}
It is now convenient to introduce a non-local field $\Phi(y) = \int G(y-z)d_z\star_z{\tilde A}$ which effectively plays the role of a canonical momentum for $-d_2 \pi^{{\tilde A}}$,as the following relation hold:
\begin{equation}
   [-d_2 \pi^{{\tilde A}}(x),\Phi(y)]
   = -i\delta^2(x-y)Vol_\Sigma
\end{equation}

With $\Phi(x)$ is possible to construct the non-local operator
\begin{eqnarray}
    \label{R-mat}
    R = -\int_\Sigma A_0 b+\pi^c\Phi
\end{eqnarray}
satisfying the following property
\begin{eqnarray}
    \label{S-mat}
    [Q_0, R] = i S\,, 
    ~~~~  S = \int_\Sigma c\pi^c-b\pi^b+iA_0 \rho+i(d_2 \pi^{{\tilde A}})\Phi   
\end{eqnarray}
Here, S acts as a number operator, inducing a natural grading on the Hilbert space.

The intertwiner can then be computed by solving the following differential equation:
\begin{equation}
    \frac{\partial\Omega}{\partial t} = i\Omega(t)[Q_1,R]_+
\end{equation}
where 
\begin{align}
     \Gamma \equiv  [Q_1,R]_+ =& -ie\int_\Sigma(\pi^{{\phi}^\dagger}{\phi}^\dagger-\pi^\phi\phi)\Phi
\end{align}

The resulting intertwiner is
\begin{equation}
    \Omega(t) = e^{i\Gamma t}
\end{equation}
We note that the intrtwiner in the present case parallels the case of QED. In the integral the charge density $(\pi^{{\phi}^\dagger}{\phi}^\dagger-\pi^\phi\phi)$ appears and it integrated with the non-local field $\Phi(x)$. The present example will be utilized in the context of spontaneous symmetry breaking in a forthcoming publication.   

Attention now turns to the construction of the invariant field $\phi^{inv}(x)$.
To this purpose
\begin{equation}
    \phi^{inv}(x)= \Omega^\dagger(t)\phi(x)\Omega(t) 
\end{equation}
is defined. A straightforward computation shows that 
\begin{equation}
    [\Gamma,\phi(x)] = e\Phi(x)\phi(x) 
\end{equation}
so, it is readily verified that
\begin{align}
    &\phi^{inv}(x) = exp(it\,{\rm ad}_\Gamma)\phi(x) = \sum_n\frac{(it)^n}{n!}{\rm ad}_\Gamma^n \phi(x) = e^{iet\Phi(x)}\phi(x)
\end{align}
where $ad_\Gamma$ means the adjoint action of the operator $\Gamma$. 
To verify that these fields are indeed gauge-invariant, it must be shown that their BRST transformation vanishes. This will now be demonstrated:
\begin{align}
    &s\phi^{inv}(x) = [Q,\phi^{inv}(x)]= \Bigg[\int_{\Sigma} c\,  (d_2 \pi^{{\tilde A}}+ie(\pi^{{\phi}^\dagger}{\phi}^\dagger-\pi^\phi\phi)) +\pi^b\rho \;,\,\phi^{inv}(x) \Bigg]\nonumber=\\
    &\int_{\Sigma}[c(y)d_y\pi^{{\tilde A}}(y)\;,e^{iet\Phi(x)}\phi(x)] -ie[c(y)\pi^\phi(y)\phi(y)\;,e^{iet\Phi(x)}\phi(x)] = (t-1)ec(x)\phi^{inv}(x)
\end{align}
that in the case of $t=1$, it is verified.
Meanwhile, if the operator $s$ is applied to $\tilde A(x)$, the usual BRST transformations are recovered
\begin{align}
    s{\tilde A}^{inv}= \big[Q,{\tilde A}^{inv}(x)\big] = \big[Q,{\tilde A}(x)\big] = \Bigg[\int_\Sigma c(y)d_y\pi(y),{\tilde A}(x)\Bigg] = -d_xc(x)
\end{align}

\section{Chern-Simons theory}
   
Three-dimensional non-abelian Chern–Simons theory is defined by the action
\begin{eqnarray}
    \label{csA}
    S = \int_{\mathcal{M}} {\rm Tr}\left(\frac12 A d A + \frac13 A\wedge A\wedge A\right). 
\end{eqnarray}
Here $\mathcal{G}$ denotes a given gauge group, the indices $a,b,c,\dots$ label the adjoint representation, $g^{ab}$ is the Killing–Cartan metric on the Lie algebra, and $f_{abc}$ are the associated structure constants.
A 2+1 decomposition of the gauge connection $A$ is performed as $A = A_0 dt + {\tilde A}$, where 
${\tilde A}$ is the connection on the Cauchy surface $\Sigma$ obtained by 
decomposing the $3$-dimensional manifold $\mathcal{M}$ into a foliation along 
the time direction vector field $\partial_0$. 
Inserting this decomposition into the action yields 
\begin{eqnarray}
    \label{csB}
    S= \int dt\int_\Sigma \mathcal{L}_2 = \int dt\int_\Sigma {\rm Tr}\left( - \frac12 {\tilde A} \partial_0 {\tilde A} + 
     A_0  F_2\right) 
\end{eqnarray}
where $F_2 = d {\tilde A} + \frac12 {\tilde A}\wedge {\tilde A}$. 
The canonical momenta are 
\begin{eqnarray}
    \label{csC}
    \pi^1 = \frac{\partial \mathcal{L}_2}{\partial (\partial_0 {\tilde A})} =  \frac12 {\tilde A}\,, ~~~~~~~~~~~~~~~~~~
    \pi^0 = \frac{\partial \mathcal{L}_2}{\partial (\partial_0 A_0)} =0\,. 
\end{eqnarray}
where $\mathcal{L}_2$ is a $2$-form on $\Sigma$. Both $\pi^1, \pi^0$ carry an index of the adjoint representation. It follows immediately \cite{Witten:1988hf} that ${\tilde A}$ is canonically conjugate to itself, and canonical quantization leads to the equal-time commutation relations 
\begin{eqnarray}
    \label{csD}
     &&[{\tilde A}^a(\vec{x},t), {\tilde A}^b(\vec{y},t)] = i g^{ab} \delta^2(\vec{x} - \vec{y}){\rm Vol}_{\Sigma} 
\end{eqnarray}
where ${\rm Vol}_{\Sigma}$ is the volume 2-form on $\Sigma$ and ${\tilde A}^a  =A^a_\mu dx^\mu$. 
The Hamiltonian is readily obtained and takes the form
\begin{eqnarray}
    \label{csE}
    H = - \int_\Sigma {\rm Tr}(A_0  F_2) 
\end{eqnarray}
which does not generate genuine dynamics but instead acts as a constraint.
Indeed, evaluating the time derivative of $\pi^0$ yields 
\begin{eqnarray}
    \label{csF}
    \partial_0 \pi^0 = \{\pi^0, H\} = F_2 \approx 0 
\end{eqnarray}
enforcing the vanishing of the two-dimensional curvature and thereby reproducing the Chern–Simons equations of motion.
Moreover, the time derivatives of the spatial connection $\tilde A$ are obtained as
\begin{eqnarray}
    \label{csFA}
    \partial_0 \pi^1 = \frac12 \partial_0 {\tilde A} = 
     \left\{\frac12 {\tilde A}, H\right\} = 
    \frac12 d_{{\tilde A}} A_0 
\end{eqnarray}
where $d_{{\tilde A}} A_0 = d A_0 + [{\tilde A}, A_0]$.
These remaining equations enforce the vanishing of the curvature, reproducing the covariant equations in 3D. 

In order to quantize the model, a gauge fixing must be imposed. To this end, a gauge-fixing term is added to the action \eqref{csA} 
\begin{eqnarray}
    \label{csG}
    \mathcal{L}^{g.f.} = {\rm Tr}
    \left(\rho \, d^\dagger {\tilde A} - b\, d^\dagger d_{{\tilde A}} c \right)  
\end{eqnarray}
where $(\rho, b)$ is the usual non-minimal Nakanishi-Lautrup/anti-ghost multiplet of $2$-forms, $d_{{\tilde A}} c = d c + [{\tilde A}, c]$ is the covariant derivative, and the Hodge dual refers to a choice of a $3$-dimensional metric. The choice of the gauge-fixing \eqref{csG} modifies the Hamiltonian and BRST charge. The latter can be computed by first computing the BRST current due to the gauge fixing, namely
\begin{eqnarray}
    \label{csGA}
    J^{g.f.}_3 = \star\rho\wedge \star d_{{\tilde A}} c\,, ~~~~~
    J^{g.f.}_1 = \star J^{g.f.}_3 = \star\rho \wedge d_{{\tilde A}} c
\end{eqnarray}
where the Hodge dual does not act on $\star\rho$. The 
current $J^{g.f.}_3$ is closed, while its Hodge dual, $J^{g.f.}_1$, is conserved. This contributes to a term in the BRST as follows 
\begin{eqnarray}
    \label{csGB}
   Q^{g.f.} = \int_\Sigma \rho (\partial_0 c + [A_0, c])
\end{eqnarray}
The second term generates the BRST transformation of $\rho$, as 
$[Q,\rho] = [\rho, c]$ and it also modify the BRST transformation of $b$. Nonetheless, it can be discarded by a simple redefinition of the fields $\rho$ and $b$, such that they behave as BRST doublets.  

The BRST charge takes the form
\begin{eqnarray}
    \label{csH}
    Q = \int_\Sigma {\rm Tr} \left(c F_2 + \rho \, \partial_0 c + \frac12 \partial_0 b[c,c] \right) 
    = 
    \int_\Sigma {\rm Tr} \left(c F_2 + \rho \, \pi^b - \frac12 [c,c] \pi^c \right)
\end{eqnarray}
where the commutator \eqref{csD} has been used, together with the conjugate momenta 
$\pi^c = - \partial_0 b, \pi^b = \partial_0 c$, both valued in the adjoint representation,
and the canonical equal-time anticommutation relations for the ghost fields. 
\begin{eqnarray}
    \label{csI}
    &&[\rho_a(\vec{x},t), A_0^b(\vec{y},t)] = i \delta_a^{b} \delta^2(\vec{x} - \vec{y}){\rm Vol}_{\Sigma}\,, ~~~~~
    \\
    &&[b_a(\vec{x},t), \pi^{b,b}(\vec{y},t)]_+ = \delta^{a}_b \delta^2(\vec{x} - \vec{y}){\rm Vol}_{\Sigma}\,, ~~~~~~~\nonumber \\
    &&[c^a(\vec{x},t),\pi^c_b(\vec{y},t)]_+ = \delta^{a}_b \delta^2(\vec{x} - \vec{y}){\rm Vol}_{\Sigma} \nonumber 
\end{eqnarray}
The BRST is nilpotent due to the first class algebra $F^a_2 = {\mathcal F}^a {\rm Vol}_\Sigma$
\begin{eqnarray}
    \label{csIA}
    [{\mathcal F}^a(\vec{x},t), {\mathcal F}^b(\vec{y},t)] = i f^{ab}_{~~c} \, {\mathcal F}^c(\vec{x},t)\delta^2(\vec{x}-\vec{y}) 
\end{eqnarray}
The BRST charge admits a decomposition into two contributions,
\begin{eqnarray}
    \label{csIAA}
    Q_0 = \int_\Sigma {\rm Tr} \left(c \, d {\tilde A} + \rho  \pi^b\right) \,, ~~~~~
    Q_{nl} = \frac12 \int_\Sigma {\rm Tr} \left(c {\tilde A}\wedge {\tilde A} - \pi^c [c,c]\right)  
\end{eqnarray} 
separating the quadratic part from the non-linear contribution to the BRST charge $Q$. These terms satisfy
\begin{eqnarray}
    \label{csIAB}
Q_0^2 =0\,,~~~~~~ [Q_0, Q_{nl}]_+=0\,, ~~~~~~[Q_{nl}, Q_{nl}]_+=0 \,. 
\end{eqnarray}
The first relation is almost trivial to check since the two pieces of the equation do not interact. The second relation is already meaningful since we have 
\begin{eqnarray}
    \label{csIB}
    [Q_0, Q_{nl} ]_+ = f_{abc} \int_\Sigma \left(c^a d c^b {\tilde A}^c - \frac12 d {\tilde A}^a c^b c^c \right) = 
    f_{abc} \int_\Sigma \left(- \frac12 d(c^a c^b) {\tilde A}^c - \frac12 d {\tilde A}^a c^b c^c \right) =0    
\end{eqnarray}
The last relation can be established in an analogous manner. 

\subsection{Intertwiner in Canonical Quantization}

The $R$ operator can now be constructed as 
\begin{eqnarray}
    \label{csL}
    R = \int_\Sigma {\rm Tr} \left( 
    b\, A_0 - \pi^c \int_\Sigma G(x-y) \, d\star {\tilde A}\right) 
\end{eqnarray}
where the first integral is taken over the $x$-coordinates, while the second is over the coordinates $y$. The Green's function $G(x-y)$ satisfies 
$\nabla^2 G(x-y) = -\delta^2(x-y)$. Evaluating the commutator between $Q_0$ and $R$ yields
\begin{eqnarray}
    \label{csM}
    [Q_0, R]_{+} = i S\,, 
    ~~~~  S = \int_\Sigma 
    {\rm Tr} \left( - b \pi^b + c \pi^c - i \rho A_0 -   i \, dA \int_\Sigma G(x-y) \, d\star A \right) 
\end{eqnarray}
where $S$ is the counting operator assigning charge $1$ to the fields $c, \rho, \pi^b$ and charges $-1$ to $b, \pi^c, A_0$. Furthermore, the fields $F^a_2 = d A^a_1$ (transverse mode) and $ d\star A^a_1$ (longitudinal mode) carry respectively $S$-charge $-1$ and $+1$. This mismatch between the charges of longitudinal and transverse modes complicates the analysis of the intertwiner. A significant simplification arises in the case of holomorphic quantization, as will be shown in the next subsection. 
According to these assignments, the $Q_0$ and $R$ have vanishing charge, and naturally, so does $S$. 

The gauge field can be decomposed into longitudinal and transverse components as 
\begin{eqnarray}
    \label{decA}
    A^a_\mu &=& \partial_\mu L^a + \epsilon_{\mu\nu} \partial^\nu T^a\,, ~~~~~~~
    \nonumber \\
    L^a &=& \frac{1}{\partial^2} \partial^\mu A_\mu^a = \int_\Sigma d^2 y \, 
    G(x-y) \partial^\rho A^a_\rho(y) \,, ~~~~~
    \nonumber \\
    T^a &=& \frac{1}{\partial^2} F^a = \int_\Sigma d^2 y \, 
    G(x-y) \epsilon^{\sigma\rho} \partial_\sigma A^a_\rho(y) \,, 
\end{eqnarray}
with the canonical commutation relations 
\begin{eqnarray}
    \label{decB}
 [L^a(x), F^b(y)] = i g^{ab} \delta^2(x-y)\,, ~~~~~ 
    [L^a(x), T^b(y)] = i g^{ab} G(x-y). 
\end{eqnarray}
The action of the linear component of the BRST charge $Q_0$ over the following fields are
\begin{eqnarray}
    \label{decBA}
    [Q_0, L^a] = c^a\,, ~~~~
    [Q_0, c^a] =0\,, ~~~~
    [Q_0, \pi^{c,a}] = F^a\,, ~~~~  
    [Q_0, F^a] =0\,. 
\end{eqnarray}
These relations show that the longitudinal mode is not BRST invariant, while the transverse mode is BRST exact. Consequently, in the Chern-Simons theory, there are no local elements in the cohomology. This seems to indicate that the cohomology is trivial and there is no physical content of the theory. This is a well-known situation: introducing non-local fields, the cohomology can be trivialized. Nonetheless, a more careful analysis of where the local BRST cohomology sits. 

One can easily check that the BRST transformations \eqref{decBA} show that the following expressions are non-trivial cohomology representatives in the space of compact support 
distributions  
\begin{eqnarray}
    \label{decBB}
    Y^a = c^a \delta(L^a) = \Big[Q_0, \Theta(L^a) \Big]_+ \,,
\end{eqnarray}
where $\Theta(L^a)$ is the Heaviside theta function (the latter is not a compact-support distribution), and the index $a$ of the adjoint representation of the gauge group is not summed. 
Note that both $c^a$ and $L^a$ carry 
charge $+1$, but $\delta(L^a)$ carries negative charge $-1$, therefore $Y^a$ has zero charge, as it should for being in the cohomology of $Q_0$.

Any product of $Y^a$ is again any element of cohomology. Since both $c^a$ and $\delta(L^a)$ are anticommuting, the maximum number of $Y^a$ is equal to the rank $r$ of the gauge group $G$. Then, we have that the highest ghost number cohomology is given by 
\begin{eqnarray}
    \label{decBC}
    Y = \prod_{i =1}^r c^{a_i} \delta(L^{a_i}) 
\end{eqnarray}
The form of these cohomologies is of the form of the puncture operator \cite{Distler:1990ea} in topological strings and the picture changing operator in string theory and supergravity \cite{Friedan:1985ge, Berkovits:2004px,Castellani:2014goa}. Furthermore, it is easy to check that $ dY^a = [Q_0, d L^a \delta(L^a)]$. 
However, to reproduce the Chern-Simons local BRST cohomology (see, for example, \cite{Barnich:2000zw}\footnote{In the case of $G= SU(N)$, the cohomology is spanned by ghost polynomials as follows $Tr(c^{2k-1})$ with $k=3,5,\dots, 2N -1$. 
The Hilbert polynomial representing the cohomology 
$P_{SU(N)}(t) = \prod_{k=3}^{2N-1} (1 - t^k)$.}) which coincides with the Chevalley-Eilenberg classes of the gauge group $G$, namely $H^n_{loc}(Q) = H^n(\mathfrak{g}, \mathbb{R})$ with $\mathfrak{g}$ the Lie algebra of the gauge group G, we have to select among the products of $Y^a$ those that are invariant under the gauge transformations. Indeed, by 
defining the new differential operator 
\begin{eqnarray}
    \label{symmGEA}
    {\mathcal W}^a = [Q_0, f^{a}_{~bc} \int_\Sigma \left(b^b A^c_0 -\pi^{c,b} L^c\right) ]_+
\end{eqnarray}
we generate the rigid rotations in the adjoint representation of the gauge group acting on the fields $c^a, L^a, \pi^a, F^a$ and on $b^a, A^a_0$ and their conjugated momenta (the structure parallels the form of $R$ in \eqref{csL}). Note that it is $Q_0$-exact (in the same way as $S$ in the commutator between $Q_0$ and $R$), and this implies that the cohomology is concentrated in the invariant subspace of the rigid transformations.  Any non-invariant quantity is BRST exact. Regarding the expression \eqref{decBC}, it is easy to check its invariance. However, we can select other products with less $Y^a$ which are invariant expressions 
of the form 
\begin{eqnarray}
    \omega^{(3)} = f_{abc} c^a c^b c^c  \left( \delta(L^a) \delta(L^b) \delta(L^c) + g_{(rs), [abc]}L^r L^s \delta'(L^a) \delta'(L^b) \delta(L^c) + \dots\right) 
\end{eqnarray}
where $g_{(rs), [abc]}$ are suitable coefficients fixed by the gauge group and by the cocycle $\omega^{(3)}$. As discussed in \cite{Catenacci:2020ybi,Cremonini:2022vgz} the additional pieces $L^r L^s \delta'(L^a) \delta'(L^b) \delta(L^c) + \dots$ are needed to guarantee the invariance of the $\omega^{(3)}$. 
It is easy to see how the transformation rule for the ghost field $c^a$, but it is less easy to build the invariant expressions in terms of Dirac delta functions are their derivatives. In paper \cite{Cremonini:2022vgz}, the full procedure is given using group theoretical methods, and an explicit realization is given in terms of Bessel functions. In the case of $SU(2)$ gauge group, the only cocycle is 
\begin{eqnarray}
    \label{su2}
    \omega^{(3)} = c^+ c^- c^0 \delta(L^+) \delta(L^-) \delta(L^0)
\end{eqnarray}
where we used the conventional notation for $SU(2)$ generator in the adjoint representation.\footnote{In the case of $SU(2)$ we can build two invariant expressions in terms of Dirac deltas $\delta_{INV}(L^0) = e^{-\frac12 L^+ L^- \partial^2_0}\delta(L^0)$
where $\partial_0 = \frac{\partial}{\partial L^0}$ and 
$\delta_{INV}(L^+, L^-) = e^{-\frac12 (L^0)^2 \partial_+ \partial_-}\delta(L^+) \delta(L^-)$
where $\partial_\pm = \frac{\partial}{\partial L^\pm}$. Multiplying those invariant expressions, we have $\delta_{INV}(L^0) \delta_{INV}(L^+, L^-) = \delta(L^+) \delta(L^-) \delta(L^0)$ which is invariant since the product of three delta transforms as the inverse of the determinant of the $SU(2)$, which is equal to one. 

}

At last, we have to notice that we can also build other elements of the cohomology by 
\begin{eqnarray}
    \label{ZA}
    Z^a = \pi^{c,a} \delta(F^a) 
\end{eqnarray}
and their (invariant) products. In the case of Chern-Simons, on-shell, $F^a \sim 0$; therefore, we do not need them. For other models, they might play a role in the construction of observables.\footnote{The operators $Y^a$ in \eqref{decBB} and $Z^a$ are \eqref{ZA} correspond to the well-known {\it Picture Changing Operators} of string theory, which are useful in the context of super-geometrical formulation of supergravity \cite{Friedan:1985ge, Berkovits:2004px,Castellani:2014goa}}.In addition, there are also non-local observables such as Wilson lines and correlation functions.  A complete analysis of the observables will be postponed in a subsequent publication. 

A straightforward computation gives
\begin{eqnarray}
    \label{commA}
[S, A^a_\mu] =  \partial_\mu L^a - \epsilon_{\mu\nu} \partial^\nu T^a
\end{eqnarray}
demonstrating that the longitudinal mode 
$L^a$ carries positive charge, whereas the transverse mode $T^a$ carries negative charge. 

Let us now consider the following expressions
\begin{eqnarray}
    \label{expA}
    ||A||^2 &&= \frac12 \int_\Sigma g_{ab} A^a \wedge \star A^b = 
    \frac12 g_{ab} \int d^2x g^{\mu\nu} 
    A^a_\mu A^b_\nu
    \nonumber \\
    &&= \frac12 
    \int_\Sigma d^2x g_{ab} \left( 
    \frac{\partial_\mu}{\partial^2} (\partial\cdot A^a) +
    \frac{\epsilon_{\mu\nu}\partial^\nu}{\partial^2}F^a
    \right) g^{\mu\rho}
    \left( 
    \frac{\partial_\rho}{\partial^2} (\partial\cdot A^b) +
\frac{\epsilon_{\rho\sigma}\partial^\sigma}{\partial^2}F^b
    \right) \nonumber 
    \\
&&=  \frac12 \int_\Sigma d^2x g_{ab} \left( 
   T^a  \frac{1}{\partial^2} T^b
   -
    L^a \frac{1}{\partial^2} L^b  
    \right)  
\end{eqnarray}
This expression shows that the longitudinal modes have a negative scalar product and are therefore unphysical, whereas the transverse modes have a positive scalar product. Transverse and longitudinal modes are mutually orthogonal with respect to this scalar product. According to the charge assignment, the first term carries charge $-2$ while the second term carries charge $+2$. 

Focus is placed on the first term in $Q_{nl}$
\begin{eqnarray}
    \label{expB}
    \frac12 \int_\Sigma f_{abc} c^a 
    A^b \wedge A^c &=& \frac12 \int d^2x\, f_{abc} c^a \epsilon^{\mu\nu} A^b_\mu A^c_\nu
     \\
     &=& \frac12 \int d^2x f_{abc} c^a \Big(  \epsilon^{\mu\nu}  
     \partial_\mu L^b \partial_\nu L^c  - 
     \epsilon^{\mu\nu}  \partial_\mu T^b \partial_\nu T^c  
     + 2 \partial^\mu L^b \partial_\mu T^c 
    \Big) 
    \nonumber
\end{eqnarray}
Computing the variation of this term under $Q_0$, one finds that only the last term leads to a non-vanishing contribution which precisely cancels $\frac12\int f_{abc} \pi^{c, a} c^b c^c$. According to this decomposition, the non-linear piece of the BRST $Q_{nl}$ contains pieces with charge $+1, +3, -1$; consequently, the BRST charge admits the following decomposition
\begin{eqnarray}
    \label{decAB}
    Q = Q_0 + Q_1 + Q_{-1} + Q_3 
\end{eqnarray}
with the following relations among the different pieces
\begin{eqnarray}
    \label{decC}
    &&Q^2_0 + \{Q_1, Q_{-1}\} =0\,, ~~~~
    \{Q_0, Q_1\} =0\,, ~~~~~~
    \{Q_0, Q_{-1}\} =0\,, ~~~~~~
    \{Q_0, Q_3\} =0\,,\nonumber \\
    &&Q^2_1 + \{Q_{-1}, Q_3\} =0\,~~~~~~
    \{Q_1, Q_3\} =0\,, ~~~~~~\{Q_{-1}, Q_{-1}\}=0\,, ~~~~\{Q_3, Q_3\}=0\,.
\end{eqnarray}
It can be shown that the non-quadratic part of the BRST charge, $Q_{nl}$, can be expressed as $Q_0$-exact in the following way:
\begin{eqnarray}
    \label{expEZ}
       Q_{nl} = [Q_0, \Gamma]
\end{eqnarray}
where  
\begin{eqnarray}
    \label{expEA}
        \Gamma &=& \frac12\int_\Sigma {\rm Tr}
    \left( (A\wedge A - [\pi^c, c] + T \, d \star A) L \right) 
        \nonumber \\
    & = &
     \frac12\int d^2x f_{abc} \epsilon^{\mu\nu} (A^a_\mu A^b_\nu - \pi^{c,a}_{\mu\nu} c^b)(x)  
     \int d^2y \, G(x-y) \partial^\rho A_\rho^c (y) 
     \nonumber \\
     &+& \frac12 \int d^2x f_{abc} 
     T^a(x) \partial^\rho A_\rho^b(x) \int d^2y G(x-y) \partial^\sigma A^c_\sigma(y)      
\end{eqnarray}
The last term of $\Gamma$ involves $T^a$, which is non-local, and maintaining two-dimensional covariance makes the operator rather cumbersome.
Since the method of \cite{grassi-porrati} cannot be applied, it is natural to ask whether it is possible to exponentiate $\Gamma$ 
by a similarity transformation $Q = e^{-\Gamma} Q_0 e^{\Gamma}$ to get the original BRST charge. Expanding this expression gives $Q = Q_0 + [Q_0, \Gamma] + \frac12 [[Q_0, \Gamma], \Gamma] + \dots$ and, for the series to truncate at the second term, the condition $[[Q_0, \Gamma], \Gamma] = 0$ and therefore $[Q_{nl}, \Gamma] = 0$ must be satisfied.

Note that the charge $Q_{nl}$ is the (ghost-fied)-Lie algebra cohomological vector field which generates the following transformation rules 
\begin{eqnarray}
    \label{expEB}
    [Q_{nl}, c_a]_+ = \frac12 f_{abc} c^b c^c\,, ~~~~
    [Q_{nl}, \pi^{c}_a]_+ = f_{abc} \Big(\frac12 A^b_\mu \epsilon^{\mu\nu} A^c_\nu -  \pi^{c,b} c^c\Big)\,, ~~~~
    [Q_{nl}, A^a_\mu] = f_{abc} c^b A^c_\mu\,. 
\end{eqnarray}
Therefore, to check whether $[Q_{nl}, \Gamma] = 0$ we have to act on $L^a$ and $T^a$, which are non-local fields \eqref{decA} that transform in a non-linear way. 

On the other hand, if $\Gamma$ can be expressed as $[Q_{nl}, R]_+$ for some anticommuting operator $R$, as in \cite{grassi-porrati}, then the nilpotency of $Q_{nl}$ ensures the required property. In the present case, however, the situation is a little more involved because of the non-linear transformations of $L^a$. Nonetheless, it can be observed that
\begin{eqnarray}
    \label{expEC}
    \bigg[Q_0, [Q_{nl}, L^a]- \frac12 f_{abc} L^b c^c\bigg]_+ = 0\,, 
    ~~~~~\Longrightarrow ~~~~
    [Q_{nl}, L^a] =  \frac12 f_{abc} L^b c^c + [Q_0, B^a]_+
\end{eqnarray}
where $B^a$ is a non-local expression of the fields. Using the properties of $L^a$  and $T^a$ under $Q_0$, it can be shown that there is no cohomology at the given ghost number. Then, using operator $R = \int d^2x (b_a A^a_0 + \pi^c_a L^a)$, one find that 
\begin{eqnarray}
    \label{expED}
   [Q_{nl}, R]_+ = \frac12
    \int d^2x \, f_{abc} (A^b_\mu \epsilon^{\mu\nu} A^c_\nu- \pi^{c,b} c^c) L^a - \int d^2x \, \pi^{c}_a [Q_0, B^a]
    = \Gamma + \Big[Q_0, \int \pi^c_a B^a\Big]_+
\end{eqnarray}
where $B^a = \frac12 f^a_{bc} (d \star \tilde A)^b L^c$. 
Finally, it is found that the correct intertwiner can be constructed using
\begin{eqnarray}
    \label{expEF}
    \hat \Gamma = \Gamma - \Big[Q_0, \int \pi^c_a B^a\Big]
\end{eqnarray}
since it also gives $Q_{nl} = [Q_0, \hat \Gamma]$, but now $\hat \Gamma = [Q_{nl}, R]_+$, which automatically implies the vanishing of the next term in the expansion of the series. In the subsequent section, we will show how this operator is simplified in the holomorphic quantization.

\subsection{Intertwiner in Holomorphic Quantization} 

An alternative and advantageous approach to quantizing Chern–Simons theory is known as holomorphic quantization \cite{Elitzur:1989nr,Axelrod:1989xt,Nair:2016ufy}. 
It is assumed that the Cauchy surface $\Sigma$ possesses a symplectic structure and has no boundary. By introducing the decomposition in terms of complex coordinates $A^a_z$ 
and $A^a_{\bar z}$ which satisfy the commutation relations 
\begin{eqnarray}
    \label{holA}
    \left[A_z(z_1), A_{\bar z}(z_2)\right] = i g^{ab} \delta^2(z_1 - z_2) 
\end{eqnarray}
we can express the BRST charge in a simplified form as 
\begin{eqnarray}
    \label{holB}
    Q &=& \int_{\Sigma} 
    \left( c_a (\partial_z  A^a_{\bar z} - \partial_{\bar z} A^a_z + f^a_{bc}
    A^b_z A^c_{\bar z}) + \rho^a \pi^b_a + \frac12 f^a_{bc} \pi^c_a  c^b c^c
    \right) \nonumber \\
    &=& Q_0 + Q_1 + Q_2 
\end{eqnarray}
where $Q_0 = \int_{\Sigma} 
    \left( c_a \partial_z  A^a_{\bar z}  + \rho^a \pi^b_a\right)$. 
    The different pieces have been separated as follows: 
    First, it is introduced the $R$ operator
\begin{eqnarray}
    \label{holC}
    R &=& \int_\Sigma d^2z \left( 
    b_a A^a_0 + \pi^c_a(z) \int_\Sigma d^2w G(z-w) \partial_{\bar w} A^a_{w}(w) 
    \right) \nonumber \\
    &=& 
    \int_\Sigma d^2z \left( 
    b_a A^a_0 + \pi^c_a(z) \int_\Sigma d^2w \frac{1}{\bar z-\bar w} A^a_{w}(w) 
    \right) 
\end{eqnarray}
where the last integral resembles a Cauchy-like formula, but it is not a line integral. Then, the anti-commutator between $Q_0$ and $R$ is computed and one obtains
\begin{eqnarray}
    \label{holD}
    [Q_0, R]_+ = i S\,, ~~~~
    S = \int_\Sigma \left( 
    - b^a \pi^b_a + c^a \pi^c_a + 
    i \rho_a A^a_0 - i g_{ab} A^a_z A^b_{\bar z} 
    \right) 
\end{eqnarray}
since $\partial_z  \partial_{\bar z} G(z-w) = \delta^2(z-w)$. This defines counting operator $S$ which assigns $+1$ charge to $c, A_z$ and $\pi^b$, while 
$b,A_{\bar z}$ and $A^a_0$ have negative charge.
In this way, the BRST charge is consistently 
filtered into three pieces, with zero, +1, and +2 charges
\begin{eqnarray}
    \label{holE}
    Q_1 &=& \int_{\Sigma} 
    f_{abc} \left( c^a
    A^b_z A^c_{\bar z} + \frac12 \pi^{c,a}  c^b c^c
    \right) \nonumber \\
    Q_2 &=& g_{ab}\int_\Sigma c^a \partial_{\bar z} A^b_z
\end{eqnarray}
The different pieces of the BRST charge satisfy the following equations
\begin{eqnarray}
    \label{holEA}
    Q_0^2 =0\,, ~~~~~[Q_1, Q_0]_+ =0\,, ~~~~~
    [Q_2, Q_0]_+ = 0\,, ~~~~~Q_1^2 =0\,, ~~~~~
    [Q_1,Q_2]_+ = 0\,, ~~~ Q^2_2 =0\,
\end{eqnarray}
and $Q_1$ generates the rigid transformations (Chevalley-Eilenberg differential). 
The intertwiner is obtained by exponentiating the following expression 
\begin{eqnarray}
    \label{holF}
    \Gamma &=& \int_\Sigma
    f_{abc} \left( 
    A^b_z A^c_{\bar z} + \pi^{c,b}  c^c
    \right) \int_\Sigma \frac{1}{\bar z-\bar w} A^a_w(w) 
    - \int_\Sigma 
    \pi^{c,a}  \int_\Sigma \frac{1}{\bar z-\bar w} f_{abc} c^b A^{c}_w(w) \nonumber \\
    &+& \frac12 g_{ab} 
    \int_\Sigma \partial_{\bar z} A^a_z \int_\Sigma \frac{1}{\bar z - \bar w} A^b_{z}(w) \\
    &=& \int_\Sigma
    f_{abc} \left( 
    A^b_z A^c_{\bar z} + \pi^{c,b}  c^c
    \right) \int_\Sigma \frac{1}{\bar z-\bar w} A^a_w(w)
    - \int_\Sigma 
    \pi^{c,a}  \int_\Sigma \frac{1}{\bar z-\bar w} f_{abc} c^b A^{c}_w(w) \nonumber \\
    &+& \frac12 g_{ab} 
    \int d^2z d^2w \partial_{\bar z} A^a_z(z) 
    G(z-w) 
    \partial_{\bar w} A^b_{w}(w) \nonumber 
\end{eqnarray}
The normal ordering is unessential since the structure constants allow us to remove the contractions between the two fields $A_z, A_{\bar z}$ on $\Sigma$.  In the second line, the last term has been symmetrically rewritten.

\subsection{Non-Abelian Wilson Loop}

In order to construct the non-abelian Wilson line, we adopt the following contruction. We recall that non-abelian Wilson loop is defined as follows: chosen a closed contour $\gamma$ (we assume that the contour $\gamma$ is in the 2D Cauchy surface $\Sigma$ orthogonal to time direction), chosen a representation $R$ for the gauge connection $A$ we define 
\begin{eqnarray}
    \label{wilA}
    {\mathcal W}_{R,\gamma}[A] = Tr_R{\mathcal P} \, 
    {\rm exp}\left( {i \oint_\gamma A_\star}\right) 
\end{eqnarray}
where $A_\star$ is the pull back of the connection on the contour. We denote by $\tau$ the variable on the contour. As is well-known the path-ordering ${\mathcal P}$ is needed in order to have the gauge invariance of ${\mathcal W}^R_\gamma[A]$ under non-abelian gauge transformations. The path-ordering leads to some difficulties in evaluating the Wilson loop and we can adopt a different realization \cite{Dorn:1986dt, Beasley:2009mb, Alekseev:2012wc, Bianchi:2024sod} by introducing a set of anticommuting set of fields $Z$ and $\bar Z$ in representation $R$, thus we get
\begin{eqnarray}
    \label{wilB}
    {\mathcal W}_{R,\gamma}[A] = \int 
    {\mathcal D}Z {\mathcal D}\bar Z  \, 
    {\rm exp}\left[ {\int_\gamma d\tau \bar Z \left(\frac{d}{d\tau} + i A_\star\right) Z}\right] 
\end{eqnarray}
which is a particle action coupled to the pull-back of the gauge connection 
$A_\star(\tau) = A_\mu(x(\tau)) \dot x^\mu d\tau$. Integrating over the coordinates $Z, \bar Z$, it reproduces the Wilson loop above (the Green function of $d/d\tau$ on the line is the Heaviside $\Theta(\tau -\tau')$ bringing the path-ordering back form the path-integration. The quantization of this system can be done by considering at first the free part. The conjugate momentum to $Z$ is the barred variable $\bar Z$, and they satisfy the usual equal time (the worldline coordinate of the quantum mechanical system) anticommutator $\{\bar Z, Z\} =1$, the Hamiltonian-Jacobi equations are 
\begin{eqnarray}
    \label{wilC}
    \dot Z + i A_\star Z=0\,, ~~~~~
    \dot {\bar Z} - i \bar Z A_\star=0\,. 
\end{eqnarray}
where the order between $A_\star$ and $Z$ or $\bar Z$ is important. 
The particle action appearing in \eqref{wilC} is BRST invariant under 
\begin{eqnarray}
    \label{wilD}
    [Q, Z] = - i c_\star Z\,, ~~~~~~
    [Q, \bar Z] =i c_\star \bar Z\,, ~~~~~~
    [Q, A_\star] = \frac{d}{d\tau}  c_\star + [c_\star, A_\star]
\end{eqnarray}
where $c_\star$ is the pull-back of the ghost fields $c_\star = c(x(\tau))$ evaluated on the contour $\gamma$. The last equation is directly inherited from the ambient space BRST variation of the gauge field $A$. We can write the total BRST charge by adding also the contribution of the fermions $Z$ and $\bar Z$ as follows 
\begin{eqnarray}
    \label{wilE}
    Q = Q_{CS} + i \bar Z c Z
\end{eqnarray}
where $\bar Z c_\star Z = c_{\star,a} \bar Z T^a Z$ with $T^a$ a Lie-algebra generator in $R$-representation. It is easy to check the nilpotency of the BRST charge since the ghost $c_\star$ tansforms as $[Q, c_\star] = \frac12  [c_\star, c_\star]$. Nonetheless, the additional piece of the BRST charge is non-linear and, since 
the fields $Z$ and $\bar Z$ have zero charge under $S$, this term has positive charge because of ghost field. To verify that they have indeed zero charge, we observe that the 
$R$ operator does not act on those fields (in the same way as in sec. 2 with the scalar matter system coupled to D=3 Maxwell theory). 
It is easy to show that, we can add the following piece to the intertwiner $\hat\Gamma$ 
constructed in the previous section of the form 
\begin{eqnarray}
    \label{wilF}
    \Gamma_Z = i \bar Z L Z = i \bar Z T_a Z \int d^2y \, G(x(\tau) - y) \partial^\mu A_\mu^a
\end{eqnarray}
where $G(x(\tau) - y)$ is evaluated on the contour $x(\tau)$ lying on the Cauchy surface $\Sigma$. 
Finally, to reproduce the Wilson loop in the form \eqref{wilB}, we observe the follow 
relation 
\begin{eqnarray}
    \label{wilG}
    e^{- \Gamma_Z} e^{ \frac{g_{ab}}{2}\int_{{\mathcal M}_3} A^a \wedge dA^b}  
     e^{\Gamma_Z} =  ^{ \frac{g_{ab}}{2}\int_{{\mathcal M}_3} A^a \wedge dA^b + 
     i \int_{\gamma} \bar Z A_\star  Z}  
\end{eqnarray}
where $Q_0$-invariant abelian part of the action $\frac12 g_{ab}\int_{{\mathcal M}_3} A^a \wedge dA^b$ is shifted to produce the coupling between the gauge fields $A_\star$ and the fermion fields $Z, \bar Z$. We used the commutation relation 
\begin{eqnarray}
    \label{wilH}
    \left[ \Gamma_Z, \frac{g_{ab}}{2}\int_{{\mathcal M}_3} A^a \wedge dA^b\right] = 
    i \int_{{\mathcal M}_3} i \bar Z T_c Z  \left[L^c, \frac{g_{ab}}{2} A^a \wedge dA^b\right] = \int_\gamma i \bar Z T_c Z A^c_\star    
\end{eqnarray}
where we used $[L^a(x(\tau)), F^b(y)] = i g^{ab} \delta^2(x(\tau) - y) $. The factor $1/2$ is eliminated by integration by parts. Notice that having chosen the contour in the Cauchy surface, the non-local longitudinal mode $L^a(x(\tau))$ is kept at the fixed time, which allows us to perform the commutation relation in \eqref{wilH}. 

\section{BF Theory}

The three-dimensional BF theory provides an interesting framework due to the rich structure of its observables. In line with the previous sections, the observables are not discussed here; instead, the focus is on the construction of the intertwiner, whose structure is similar to three-dimensional gravity (which will be analysed in detail in the forthcoming sections). In particular, we will show how the two-steps procedure can be applied when the cosmological term is added. That term produces a negative-charge piece of the BRST charge and therefore one needs a different approach as described in sec. 2. In the present case, the complete derivation can be done explicitly.

The action is given by
\begin{eqnarray}
    \label{BFA}
    S = \int_{\mathcal{M}^3} {\Tr} B\wedge F_A\,, 
\end{eqnarray}
where $F_A = d A + \frac12[A,A]$. Both $A^a$ and $B^a$ carry the adjoint representation of the gauge group $G$. 
The classical BRST symmetry is 
\begin{eqnarray}
    \label{BFB}
    s\, A = \nabla_A c = d A + [A, c]\,, ~~~~~~
    s\, B = \nabla_A \lambda + [c, B] = d \lambda + [A, \lambda] + [B,c]\,,
\end{eqnarray}
where $c^a$ is the ghost for the $A$-gauge symmetry and $\lambda$ is the ghost for the $B$-gauge symmetry. 

In the present case, the canonical analysis is not repeated. Instead, recall that quantization proceeds in the standard way by decomposing the gauge fields $A$ and $B$ into their time components $A_0$ and $B_0$, and their spatial components ${\tilde A}$ and $\tilde B$ along the Cauchy surface $\Sigma$. With this decomposition, quantization of the fields on $\Sigma$ leads to the equal-time commutation relations 
\begin{eqnarray}
    \label{BFC}
    [{\tilde A}^a(\vec{x},t), {\tilde B}^b(\vec{y},y) ] = i g^{ab} \delta^2(\vec{x} - \vec{y}) {\rm Vol}_\Sigma
\end{eqnarray}
where $g^{ab}$ is the Killing-Cartan metric and ${\rm Vol}_\Sigma$ is the volume of the $\Sigma$ surface. The gauge fixing needed for the quantization is chosen to be 
\begin{eqnarray}
    \label{BFD}
    \nabla \star {\tilde A} = 0\,,  ~~~~~~~~ \nabla_{\tilde A} \star {\tilde B} =0
\end{eqnarray}
where $\nabla$ is the derivative with respect to the spacetime coordinates and $\nabla_{\tilde A}$ is with respect to the gauge field $\tilde A$.
A Landau–Fermi gauge fixing is adopted for both gauge fields. For the  ${\tilde B}$ field, however, this choice is modified by replacing the ordinary derivative with a covariant one, to preserve invariance under the BRST symmetry associated with the ghost  $c$. The Hodge dual is computed by adding a metric $g$ on the surface $\Sigma$. The gauge-fixing procedure requires the introduction of the Nakanishi–Lautrup auxiliary fields $\rho_A$ and $\rho_B$ together with the corresponding anti-ghost fields $b^A_a$ and $b^B_a$. Finally, the BRST charge is 
\begin{eqnarray}
    \label{BFE}
    Q = \int_\Sigma \Big( c_a \nabla_{\tilde A} {\tilde B}^a + \lambda_a F^a_{\tilde A}  + \rho^A_a \pi^{b^A, a} + 
    \rho^B_a \pi^{b^B, a} + \frac12 f_{abc} c^a c^b \pi^{c,c} + f_{abc} c^a \lambda^b \pi^{\lambda, c}
    \Big) 
\end{eqnarray}
where $\pi^{b^A}, \pi^{b^B}, \pi^c, \pi^\lambda$ are the conjugated momenta with respect to $b^A, b^B, c$ and $\lambda$. Note that the expression in the bracket is a $2$-form integrated on the Cauchy surface. Owing to the self-interaction of ${\tilde A}$ and the interaction between ${\tilde A}$ and ${\tilde B}$, the BRST charge contains several non-linear contributions. It is therefore convenient to decompose it into a quadratic part and a non-linear part as follows
\begin{eqnarray}
    \label{BFF}
    Q _0 &=& \int_\Sigma \Big( c_a \nabla {\tilde B}^a + \lambda_a \nabla {\tilde A}  + \rho^A_a \pi^{b^A, a} + 
    \rho^B_a \pi^{b^B, a} 
    \Big) \nonumber \\
    Q _1 &=& \int_\Sigma \Big( f_{abc} c^a {\tilde A}^b \wedge {\tilde B}^c + \frac12 
    f_{abc} \lambda^a {\tilde A}^b \wedge {\tilde A}^c  + \frac12 f_{abc} c^a c^b \pi^{c,c} + f_{abc} c^a \lambda^b \pi^{\lambda, c}
    \Big) 
\end{eqnarray}
where the separation of the two pieces is done according to the following counting operator $S$ that we define below. 
Firstly, the $R$ operator is introduced as
\begin{eqnarray}
    \label{BFG}
    R = \int_\Sigma  \Tr\Big( b^{ A} {A}_0 + b^B B_0 + \pi^c(x) \int_\Sigma G(x-y) \nabla_{\tilde A} \star {\tilde B}(y) + 
    \pi^\lambda(x) \int_\Sigma G(x-y) \nabla \star {\tilde A}(y)
    \Big) 
\end{eqnarray}
where the Green's function $G(x-y)$ satisfies $\nabla^2 G(x-y) = \delta^2(x-y)$.  The 
commutator between $R$ and $Q_0$ lead to the counting operator $S$ which assigns the charges 
$+1$ to $c, {\tilde A}, \pi^{b^A}, \pi^\lambda, \rho^B$ and $-1$ to $\lambda, {\tilde B}, \pi^{b^B}, \pi^c, \rho_A$. In this way, the $Q_1$ piece of the BRST charge has only positive charge. Note that ${\tilde A}$ and ${\tilde B}$ have opposite charges with respect to the commutation relation \eqref{BFC}. Finally, the intertwiner is given by $\Gamma =[Q_1, R]_+$ 
\begin{eqnarray}
    \label{BFH}
    \Gamma &=& \int_\Sigma \Big[ f_{abc} ({\tilde A}^b \wedge {\tilde B}^c + c^b \pi^{c,c} + \lambda^b \pi^{\lambda,c}) \int_\Sigma G(x-y) \nabla_{\tilde A} \star {\tilde B} 
    \nonumber \\
    &&+ 
    f_{abc} (\frac12 {{\tilde A}}^b \wedge {{\tilde A}}^c + c^b \pi^{\lambda,c}) \int_\Sigma G(x-y) \nabla \star {{\tilde A}}
 \\ &&+ \pi^{c,a} \int_\Sigma G(x-y)  \nabla_{{\tilde A}} \star f_{abc} (\lambda^b {{\tilde A}}^c + c^b {\tilde B}^c) + 
    \pi^{\lambda,a} \int_\Sigma G(x-y)  \nabla \star f_{abc} c^b {{\tilde A}}^c \Big]     \nonumber
\end{eqnarray}
Note that the last two terms are required in order to account for the linear transformations of the fields ${{\tilde A}}$ and ${\tilde B}$. Owing to the non-local nature of these expressions, they cannot cancel the contributions arising from the first two lines. 

\subsection{The Cosmological Term}

It is worth noting that BF theory admits deformations obtained by adding a term of the form
\begin{eqnarray}
    \label{BFI}
    S_{BF, cosm} = \frac{\kappa}{3!} \int_{\mathcal{M}^3} \Tr(B\wedge B\wedge B)\,.
\end{eqnarray}
where $\kappa$ is a constant parameter. 
If the gauge group $G$ is replace with the $GL(3)$ and the manifold ${\mathcal M}^3$ the frame bundle where the soldering $\theta$ is a 1-form with values in $\mathbb{R}^3$ associated to the fundamental representation of $GL(3)$, the $B^a$ field represents the dreinbein of gravity in $D=3$ expressed in local coordinates. In the following, we turn exactly to $D=3$ gravity exporting the present results. We leave to future publications the analysis of observables of BF theories along the lines of \cite{Cattaneo:1995tw,Cattaneo:2025xvk}.
Here, it is interesting to analyze the intertwiner construction. 

The contribution to the BRST charge $Q$ of the cosmological term is given by 
\begin{eqnarray}
    \label{CCA}
    Q_{-3} = \kappa \int_\Sigma {\rm Tr}
    \left( \frac{\lambda}{2}[B,B] + \frac12 \pi^c[\lambda, \lambda] \right) 
\end{eqnarray}
which carries a negative charge. Nonetheless, according to the analysis in sec. we can reabsorb into an intertwiner by a two step procedure. First, we compute the commutator with the $Q = Q_0 + Q_1$, and we find 
\begin{eqnarray}
    \label{CCB}
    [Q, Q_{-3}]_+ = 0
\end{eqnarray}

However, we can adopt a easier framework. 
Indeed we can construct the following two combinations 
\begin{eqnarray}
    \label{CCC}
    \mathcal{A}^\pm = \frac12(A \pm \kappa B) \,, ~~~~~~
    c^\pm = \frac12(c \pm \kappa \lambda) \,,
\end{eqnarray}
The total action turns out to be the sum of two Chern-Simons theories. For them we can adopt the holomorphic quantization method and therefore we can build the intertwiner along the strategy depicted in sec. 4.2.

\section{Gravity}

\subsection{2+1 Chern-Simons Gravity in Canonical Quantization}

In the present section, we analyse the case of three-dimensional gravity. This is a rather simple model, but it has some of the details needed for gravity in higher dimensions. 
We start with the dreinbein-spin connection formulation of $3D$ gravity, given by the action 
\begin{eqnarray}
    \label{graA}
    S = \int_{\mathcal{M}_3}  \epsilon_{abc} e^a \wedge R^{bc}(\omega) 
\end{eqnarray}
where $e^a$ is the dreibein $1$-form and $\omega^a$ the $1$-form spin-connection. The indices $a,b,c$ are the tangent space indices.  
The equations of motion are 
\begin{eqnarray}
    \label{graB}
    R^{ab}(\omega)  \equiv 
    d \omega^{ab} + \omega^{ac} \wedge \omega_{c}^{~~b}
    =0\,, ~~~~~~
    T^a(e, \omega) \equiv \nabla e^a = d e^a + \omega^{ab} \wedge e_b =0\,. 
\end{eqnarray}
Consistency implies that $\nabla R^{ab} =0$ and $R^{ab} \wedge e_b =\nabla T^a$, which are the Bianchi identities. Solving the tortionless condition for $\omega^{ab}$ in terms of $e^{a}$, Einstein’s equations are recovered from the curvature equation. 
The theory admits two gauge symmetries, whose BRST transformations read
\begin{eqnarray}
\label{graBA}
s\, e^a = d c^a + \omega^{ab} c_b + \lambda^{ab} e_b = \nabla c^a + \lambda^{ab} e_b\,, ~~~~
s\, \omega^{ab} = d \lambda^{ab} + \omega^{ac} \lambda_c^{~b} -  \lambda^{ac} \omega_c^{~b} \,, 
\end{eqnarray}
where $c^a$ and $\lambda^{ab}$ are the ghosts associated with diffeomorphisms and local Lorentz transformations, respectively. Their BRST variations are 
\begin{eqnarray}
    \label{graBB}
    s\, c^a = \lambda^{ab} c_b\,, ~~~~~~~
    s\, \omega^{ab} = \lambda^{ac} \lambda_{c}^{~b}
\end{eqnarray}
while the BRST variation of $R^{ab}$ and $T^a$ are proportional to the equations of motion. 

Performing the followwing $2+1$ decomposition,
\begin{eqnarray}
    \label{graC}
    e^a = e^a_0 \,dt + \hat e^a \,, 
    ~~~~~~
    \omega^{ab} = \omega^{ab}_0 dt + \hat\omega^{ab}
\end{eqnarray}
the action becomes
\begin{eqnarray}
    \label{graD}
    S = \int dt \int_{\Sigma_2} 
    \left( - \epsilon_{abc} \hat e^a \wedge\partial_0 \hat \omega^{bc} + 
    \epsilon_{abc} \hat e^a_0 
    R^{bc}(\hat \omega) -  \epsilon_{abc} \omega^{ab}_0
    \hat\nabla \hat e^c 
    \right) 
\end{eqnarray}

The canonical momenta are
\begin{align}
    \label{graE}
    &\pi_a^0 \simeq 0\,, ~~~~
    &\pi_{ab}^0 \simeq 0\,,\nonumber  ~~~~\\
    &\pi_a \simeq 0\,, ~~~~~~
    & \pi_{ab}-\e_{abc}\hat{e}^c \simeq 0\,, 
\end{align}
which constitute primary constraints. The first set (first line of \eqref{graE}) is first class, while the second set is second class, with the Poisson brackets given by
    \begin{eqnarray}
    \label{graF}
    \{\pi_a, \pi_{bc} - \epsilon_{bcd} \hat e^d \} = \epsilon_{abc} {\rm Vol}_\Sigma \, \delta^2(x -y) \,, 
\end{eqnarray}
where ${\rm Vol}_\Sigma$ is the two-dimensional volume form, and canonical brackets have been used. 
The second-class constraints are imposed strongly, eliminating $\pi_{ab}$ in favour of $\hat e^a$; the remaining conjugate variables are therefore $\hat e^a$ and $\hat \omega^{ab}$ with the commutator 
\begin{eqnarray}
   \label{graFA}
    [\hat \omega^{ab}(t,\vec{x}),  
    \hat e^c(t,\vec{y}) ] = \frac{i}{2} \epsilon^{abc}\, {\rm Vol}_\Sigma \, \delta^2(\vec{x} -\vec{y}) \,. 
\end{eqnarray}


The Hamiltonian takes the form
    \begin{eqnarray}
    \label{graG}
    H = - \int_{\Sigma_2} 
    \left( 
    \epsilon_{abc} \, e^a_0 
    R^{bc}(\hat \omega) - \epsilon_{abc} \, \omega^{ab}_0
    \hat T^c 
    \right) 
\end{eqnarray}
The time preservation of the first-class constraints yields
    \begin{eqnarray}
    \label{graH}
    \dot \pi^0_a = -\frac{\partial H}{\partial e^a_0} =  \e_{abc}R^{bc}(\hat \omega)  \simeq 0 \,, ~~~~~
    \dot \pi^0_{ab} = 
    -\frac{\partial H}{\partial \hat \omega^{ab}_0} = 
    -\e_{abc}\hat T^c \simeq 0
\end{eqnarray}
corresponding to the spatial components of the equations of motion. The remaining equations follow from the Hamiltonian equations, for instance, varying the Hamiltonian with respect to $\pi_{ab}$ yields 
\begin{eqnarray}
    \label{graL}
    \partial_0 \hat \omega^{ab} = -  \frac{\partial H}{\partial \pi_{ab}} = \hat \nabla \omega^{ab}_0
\end{eqnarray}

In order to quantize the system, gauge fixing and the associated ghost sector must be introduced. The resulting BRST charge takes the form
\begin{eqnarray}
    \label{bbA}
    Q = \int_\Sigma \left( \epsilon_{abc} c^a R^{bc}(\hat \omega) + 
    \epsilon_{abc} \lambda^{ab} T^c(\hat e, \hat \omega) 
    + \rho_a \pi^{b,a} + \rho_{ab} \pi^{b,ab} + 
    \lambda^{ac} c_c b_a + \frac12 \lambda^{ac}  \lambda_{c}^{~b} b_{ab}
    \right) 
\end{eqnarray}
where $(\rho_a, b_a)$ and $(\rho_{ab}, b_{ab})$ are the Nakanishi-Lautrup $2$-form multiplets. As for the Chern-Simons theory discussed above, the BRST charge can be decomposed into its quadratic and higher-order parts,
\begin{eqnarray}
\label{bbB}
Q_0 &=& \int_\Sigma \left( \epsilon_{abc} c^a d \hat \omega^{bc} + 
    \epsilon_{abc} \lambda^{ab} d \hat e^c 
    + \rho_a  \pi^{b,a} + \rho_{ab}  \pi^{b,ab}
    \right) \,, \nonumber \\
Q_1 &=& \int_\Sigma \left( \epsilon_{abc} c^a \hat\omega^{bd} \wedge \hat\omega_{d}^{~~c} + 
    \epsilon_{abc} \lambda^{ab} \hat \omega^{cd} \wedge \hat e_d 
    + 
    \lambda^{ac} c_c b_a + \lambda^{ac}  \lambda_{c}^{~b} b_{ab}
    \right) 
\end{eqnarray}
Introducing the non-local fields 
\begin{eqnarray}
    \label{bbC}
    \Phi^a(x) = \int_\Sigma G(x-y) \partial^\rho \hat e^a_\rho(y)\,, ~~~~~~
     \Phi^{ab}(x) = \int_\Sigma G(x-y) \partial^\rho \hat \omega^{ab}_\rho(y)\,, 
\end{eqnarray}
the following commutation relations are obtained:
\begin{eqnarray}
\label{bbD}
&&    \left[\partial_{[\mu} \hat\omega^{ab}_{\nu]}(\vec{x},t),  \Phi^c(\vec{y},t) \right]
 = \frac i2 \epsilon^{abc} \epsilon_{\mu\nu} \delta^2(\vec{x}- \vec{y}) \nonumber \\
&&    \left[\partial_{[\mu} \hat e^{a}_{\nu]}(\vec{x},t),  \Phi^{bc}(\vec{y},t) \right]
 =  - \frac i2 \epsilon^{abc} \epsilon_{\mu\nu} \delta^2(\vec{x}- \vec{y}) 
\end{eqnarray}
and by using the curvatures $R^{ab}_0 = d \hat \omega^{ab}$ and 
$T^{a}_0 = d \hat e^{a}$
\begin{eqnarray}
\label{bbE}
&&    \left[R^{ab}_0(\vec{x},t),  \Phi^c(\vec{y},t) \right]
 = i \epsilon^{abc}{\rm Vol}_\Sigma \delta^2(\vec{x}- \vec{y}) \nonumber \\
&&    \left[T^{a}_0(\vec{x},t),  \Phi^{bc}(\vec{y},t) \right]
 = i \epsilon^{abc} {\rm Vol}_\Sigma \delta^2(\vec{x}- \vec{y}) 
\end{eqnarray}

The action of $Q_0$-variation on the non-local fields $\Phi^a, \Phi^{ab}$ is readily computed, 
\begin{eqnarray}
    \label{bbF}
    [Q_0, \Phi^a(x)] =   \int_\Sigma G(x-y) \partial^\rho \partial_\rho c^a = 
    c^a\,, ~~~~~~
    [Q_0, \Phi^{ab}(x)] = \int_\Sigma G(x-y) \partial^\rho \partial_\rho \lambda^{ab} = 
    \lambda^{ab}\,, ~~~~~~
\end{eqnarray}
Further commutators yield
\begin{eqnarray}
    \label{bbG}
    \left[\omega_\mu^{ab}(x), \Phi^c(y) \right] = i \epsilon^{abc} 
    \epsilon_{\mu\nu} \partial^\nu G(x-y) \,, ~~~~~
    \left[e_\mu^{a}(x), \Phi^{bc}(y) \right] =  - i  \epsilon^{abc} 
    \epsilon_{\mu\nu} \partial^\nu G(x-y) \,, ~~~~~
\end{eqnarray}
These ingredients allow the construction of the non-local operator $R$ entering the intertwiner,
\begin{eqnarray}
    \label{bbGA}
    R = \int_\Sigma \Big( b_a e^a_0 + b_{ab} \omega^{ab}_0 + 
    \Phi^a \pi_{c,a}  + 
     \Phi^{ab} \pi_{c,ab} \Big).
\end{eqnarray}
Here, the two fields $ e^a_0, \omega^{ab}_0$ denote the temporal components of the dreibein and the spin-connection, whose canonical commutation relations
\begin{eqnarray}
    \label{bbGB}
    [\rho_a(x,t), e^b_0(y,t)] = i \delta_a^b \delta(x-y)\,, ~~~~
    [\rho_{ab}(x,t), \omega^{cd}_0(y,t)] = i \delta_{ab}^{cd} \delta(x-y)\,, ~~~~
\end{eqnarray}
follow from the chosen gauge-fixing conditions.
Then, the following commutator  
\begin{align}
    \Gamma &= [Q_1,R]_+ = - \int_\Sigma \Phi^a\bigg(\epsilon_{abc}\hat\omega^{bd}\wedge\omega_d^{\;c}+\lambda^b_{\;a}b_b \bigg) + \Phi^{ab}\bigg(\e_{abc}\hat{\omega}^{cd}\wedge\hat e_d+2\lambda_{\;b}^{c}b_{ac} +c_bb_a\bigg) +\\&i\int_{\Sigma}\int_{\Sigma}\bigg(2 \pi_{c,e}(y)\Big(\lambda^{de}(x)\hat e_d(x)-c^a(c)\hat\omega^e_{\;a}(x)\Big) +\pi_{c,ef}(y)\epsilon_{abc}\epsilon^{def}\lambda^{ab}(x)\hat\omega^c_{\;d}(x)\bigg)\epsilon_{\mu\nu}\partial^\nu_xG(x-y)dx^\mu. \nonumber 
\end{align}
The intertwiner is therefore given by
\begin{equation}
    \Omega(t) = e^{i\Gamma t}
\end{equation}
evaluated in $t=1$.

\subsection{2+1 Chern-Simons Gravity in the Time Gauge}

To describe $2+1$-dimensional Hamiltonian gravity in a manner that does not depend on the choice of a particular Cauchy surface, and within the commonly used time-gauge formalism, the spacetime is assumed to be globally hyperbolic. An orthonormal time-like vector with flat indices, $n^I$ ($I= \{0,1,2\}$), is then introduced. This vector is orthogonal to all quantities defined on the Cauchy surfaces and is normalized to $-1$.

The extrinsic curvature is defined as the projection onto the Cauchy surface (parallel component) of the covariant derivative of $n^I$:
\begin{equation}
    \label{extrinsic_curvature_definition}
    K^I = ^\parallel\!\!\nabla n^I = ^\parallel\!\!(\omega^I_{\; J}n^J)
\end{equation}

The vector $n^I$ is related to its coordinate-basis counterpart $n^\mu = (1/N,-N^i/N)$ through
\begin{equation}
    \label{tangent_flat_vector_relation}
    n^I = n^\mu e^I_\mu
\end{equation}
The tetrad is decomposed into components orthogonal and parallel to the Cauchy surface. 
In particular, it holds that $e^\mu_\perp = n^\mu$ and that the inner product of $e_\perp$ and the parallel tetrad $e^a$ (with $a=1,2$) with respect to the Cauchy surface is zero.
Before proceeding, the time gauge is fixed by choosing
\begin{equation}
    \label{time_gauge_tangent_flat_vector}
    n^I = (1,0,0)
\end{equation}
With this choice, the tetrads take the form
 \begin{eqnarray}
 \label{tedA}
        e^\perp &=& Ndt \nonumber \\
        e^a &=& N^adt+\tilde e^a \nonumber \\
        e_\perp &=& \frac{1}{N}(\partial_0-N^i\partial_i)\\
        e_a &=& \tilde e_a \nonumber 
 \end{eqnarray} 
where $N^a$ is related to the shift vector as $N^a = e^a_iN^i$ (with $i_1,2$) and $\tilde e^a$ is the tetrad that defines the metric of the Cauchy surface.

The spin connection is similarly decomposed into parallel and orthogonal components,
\begin{equation}
    \label{spin_connection_splitting}
    \omega^{IJ} =  \;\tilde\!\!\omega^{IJ} + e^\perp \omega^{IJ}_\perp
\end{equation}
where $\omega^{IJ}_\perp$ is given by the inner product between $e_\perp$ and $\omega^{IJ}$:
\begin{equation}
    \label{orto_spin_connection}
    \omega^{IJ}_\perp = \iota_{e_\perp}\omega^{IJ} = \frac{1}{N}(\omega^{IJ}_t-N^i\omega^{IJ}_i)
\end{equation}
and the parallel component is obtained from the following relation
\begin{equation}
    \label{parallel_spin_connection}
    \tilde\omega^{IJ} = \omega^{IJ} - e^\perp\omega^{IJ}_\perp = \omega^{IJ}_i(dx^i+N^idt).
\end{equation}
In this framework, the extrinsic curvature becomes
\begin{equation}
    \label{time_gauge_extrinsic_curvature}
    K^a = ^\parallel\!\!(\omega^a_{\;0})
\end{equation}
Decomposing the Riemann curvature into components parallel and orthogonal to the Cauchy surface yields
\begin{equation}
    \label{riemann_splitting}
    \begin{split}
            R^{IJ} &= \tilde R^{IJ} + e^\perp\wedge R^{IJ}_\perp\\ &={d_2} \tilde\omega^{IJ} + {\tilde\omega^{I}_{\;K}\wedge} \tilde\omega^{KJ} + e^\perp\wedge \Big[\partial_\perp(\tilde\omega^{IJ})-d_2(\omega^{IJ}_\perp)+ { \omega^{I}_{\perp\;K} }  \tilde\omega^{KJ}- \tilde\omega^{I}_K \omega^{KJ}_{\perp\;}\Big] 
    \end{split}
\end{equation}
This decomposition allows the Lagrangian density to be defined. Prior to gauge fixing, the action is
\begin{equation}
    \label{EH_action}
    \mathcal{S} = \int_{\mathcal{M}^{3}}\epsilon_{IJK}e^I\wedge R^{JK}
\end{equation}
In the present case, the gauge choice \eqref{tedA} alone is not sufficient, and an additional gauge fixing must be imposed. In particular, the condition $e^0_\parallel = 0$ affects the equations of motion. One finds the Hamilton–Jacobi equation 
\begin{equation}
    \label{Hamilton-Jacobi_Gauge_fix}
    \dot \pi_{e^0_\parallel} = -\frac{\partial \mathcal{H}}{\partial e^0_\parallel} = 0
\end{equation}
which consistently enforces the gauge-fixing condition. \eqref{Hamilton-Jacobi_Gauge_fix}. 
This is achieved by using the gauge spinor $\Psi$ 
\begin{equation}
    \label{Gauge_Spinor}
    \Psi = \int_\Sigma \overline{c}\wedge\partial_0 \pi_{e^0_\parallel} = \int_\Sigma \overline{c}\wedge\partial_0 (\epsilon_{ab}\tilde\omega^{ab})
\end{equation}
where $\overline{c}$ is the antighost 1-form related to the residual $SO(2)$ Lorentz transformation.
Applying the BRST transformations $s$ to the gauge spinor yields
\begin{equation}
    \label{BRST_gauge_spinor}
    s\Psi = \int_\Sigma B\wedge\partial_0 (\epsilon_{ab}\tilde\omega^{ab}) + \overline{c}\wedge\partial_0(\nabla_2c)
\end{equation}
where $c=\epsilon_{ab}c^{ab}$ the ghost 0-form and $B$ is the auxiliary field 1-form of Nakanishi–Lautrup.
The gauge-fixed Lagrangian is defined by
\begin{equation}
    \mathcal{L}_{gf} = \mathcal{L}_{EH}|_{e^0_\parallel=0} + \int_\Sigma B\wedge\partial_0 (\epsilon_{ab}\tilde\omega^{ab}) + \overline{c}\wedge\partial_0(\nabla_2c)
\end{equation}
The decomposition of the internal indices $I$ into $\{\perp, a\}$ is now performed in the Einstein–Hilbert Lagrangian \eqref{EH_action}. The resulting expression is 
\begin{eqnarray}
        \mathcal{L}_{EH} = \epsilon_{IJK} e^I\wedge R^{JK} = \e_{0ab}dt\wedge[N \tilde R^{ab} -2 {N^{a}}\tilde R^{0b} +2N \tilde e^a \wedge R_\perp^{0b}]
\end{eqnarray}
Where 
\begin{align}
    \tilde R^{ab} =& \;  d_2(\tilde\omega^{ab}) + {\tilde\omega^{a}_{\;c}\wedge} \tilde\omega^{cb} + {\tilde\omega^{a}_{\;0}\wedge} \tilde\omega^{0b} =\tilde R_2^{ab} - K^a \wedge K^b\\
    -2 {N^a}  \tilde R^{0b} =&-2{N^a}\Big[d_2(\tilde\omega^{0b})+\tilde\omega^{0c}{\wedge} \tilde\omega^{\;b}_c\Big] = -2{N^a}\nabla_2(K^b)\\
    2N\tilde e^a \wedge R_\perp^{0b}=& 2N\tilde e^a \wedge[\partial_\perp(\tilde\omega^{0b}) -d_2(\omega_\perp^{0b}) +{\omega^0_{\perp\;c}} \tilde\omega^{cb} - \tilde\omega^{0}_{\;c}\omega_\perp^{cb}]  \\=&  2N\tilde e^a \wedge\Big[\partial_\perp(\tilde\omega^{0b})  - \tilde\omega^{0}_{\;c}\omega_\perp^{cb}-\nabla_2(\omega_\perp^{0b})\Big] \nonumber\\
     =&2N\tilde e^a \wedge\Big[\partial_\perp(K^{b})  - K_c\omega_\perp^{cb}-\nabla_2(\omega_\perp^{0b})\Big] \nonumber
\end{align}
The Lagrangian can be written as follows: 
\begin{equation}
    \begin{split}
        {L}_{EH} &= \int \e_{ab} dt\wedge\{N\Big[\, ^\parallel\! R_2^{ab} - K^a \wedge K^b\Big] -2{N^a} \nabla_2(K^b) + 2N \tilde e^a \wedge\Big[\partial_\perp(K^{b}) -K_c\omega_\perp^{cb}  -\nabla_2(\omega_\perp^{0b})\Big]\} \\
        &= \int \e_{ab} dt\wedge\{N\Big[\, ^\parallel\! R_2^{ab} - K^a \wedge K^b\Big] -2{N^a} \nabla_2(K^b) + 2 \tilde e^a \wedge\Big[\partial_0(K^{b}) -K_c\omega_0^{cb}  -N\nabla_2(\omega_\perp^{0b})\Big]\}
    \end{split}
\end{equation}
If the manifold is sufficiently regular and boundary terms are neglected, an integration by parts on $\tilde e^a \wedge\partial_0(K^{b})$ can be performed, leading to the following Lagrangian density:
\begin{equation}
    \mathcal{L}_{EH} =  \e_{ab}\{N\Big[\,\tilde R_2^{ab} - K^a \wedge K^b\Big] -2{N^a} \nabla_2(K^b) - 2 \tilde e^a \wedge\Big[K_c\omega_0^{cb}  +N\nabla_2(\omega_\perp^{0b})\Big] -2\partial_0(\tilde e^a)\wedge K^b\}
\end{equation}
Upon integration by parts, the gauge-fixed Lagrangian density takes the form \cite{petkou}
\begin{equation}
    \begin{split}
            \mathcal{L}_{gf} =&  \e_{ab}\{N\Big[\tilde R_2^{ab} - K^a \wedge K^b\Big] -2{N^a} \nabla_2(K^b) - 2 \tilde e^a \wedge K_c\omega_0^{cb} +2 \omega_0^{0b} \nabla_2(\tilde e^a) -2\partial_0(\tilde e^a)\wedge K^b\}+\\&+B\wedge\partial_0 (\epsilon_{ab}\tilde\omega^{ab}) + \overline{c}\wedge\partial_0(\nabla_2c)
    \end{split}
\end{equation}
To construct the canonical Hamiltonian, the conjugate momentum densities are first computed:
\begin{equation}
    \label{momentum_Constrain}
    \begin{split}
        \pi_N &= \frac{\partial \mathcal{L}}{\partial\dot N} = 0, \qquad\qquad\;\;  \pi_{N,a} = \frac{\partial \mathcal{L}}{\partial\dot {N^a}} = 0, \;\;\;\Pi^0_{ab} = \frac{\partial \mathcal{L}}{\partial\dot \omega_0^{ab}} = 0,\;\;\;\Pi^0_{0a} = \frac{\partial \mathcal{L}}{\partial\dot \omega_0^{0a}} = 0\\
          \pi_a &= \frac{\partial \mathcal{L}}{\partial\dot{ \tilde e}^a} = -2\epsilon_{ab}K^b \quad\;\; \tilde\Pi_{a} = \frac{\partial \mathcal{L}}{\partial\dot K^{a}} = 0\;\;\; \\\tilde\Pi_{ab} &= \frac{\partial \mathcal{L}}{\partial\dot{ \tilde \omega}^{ab}} = -\epsilon_{ab}B\qquad\; \pi_B = \frac{\partial \mathcal{L}}{\partial\dot B} = 0 \\
        \pi_C &= \frac{\partial \mathcal{L}}{\partial\dot c} = \nabla_2 \overline{c}\qquad\qquad \pi_{\overline{C}} =  \frac{\partial \mathcal{L}}{\partial\dot{ \overline{c}}} = 0
    \end{split}
\end{equation}
It is observed that the Legendre map $\mathcal{F}_\mathcal{L}$, relating the Lagrangian variables to the Hamiltonian ones, is not invertible. In particular, the set of velocities $\dot\psi$ (where $\psi$ collectively denotes the fields of the theory) cannot be expressed as functions of the canonical momenta. As a consequence, all canonical momenta arise as primary constraints of the theory.

To analyze the constraint structure, the total Hamiltonian is introduced,
\begin{align}
        {H}_{Tot} =&  \int_\Sigma\mathcal{H}_{Canon} + \sum_i\int_\Sigma u^i\wedge\phi_i \\
        \mathcal{H}_{Canon} =& -\e_{ab}\{N\Big[\tilde R_2^{ab} - K^a \wedge K^b\Big] -2{N^a} \nabla_2(K^b) - 2 \tilde e^a \wedge K_c\omega_0^{cb} +2 \omega_0^{0b} \nabla_2(\tilde e^a) \}\\&- B\wedge\partial_0 (\epsilon_{ab}\tilde\omega^{ab}) - \nabla_2(\overline{c})\wedge\partial_0(c) \nonumber
\end{align}
Here, the index $i$ is a generic index running over all the constraints defined through the momentum equations \eqref{momentum_Constrain}.
For instance, $\phi_N := \pi_N=0$ and $\phi_a := \pi_a +2\epsilon_{ab}K^b = 0$. The $u_i$ are n-form Lagrange multipliers associated with constraints \eqref{momentum_Constrain}.
Some of these Lagrange multipliers also carry a direct physical interpretation, as they partially restore the invertibility of the Legendre map $\mathcal{F}_\mathcal{L}$. This becomes explicit in those cases where
\begin{equation}
    \label{time_derivative_equation}
    \dot\phi_i = \{\phi_i,{H}_{Tot}\} = u_i
\end{equation}
a feature that will play an important role in the consistency analysis.

The Dirac consistency algorithm is now applied to all constraints. The only non-vanishing Poisson brackets among the momenta are
\begin{eqnarray}
    \{\phi_a,\tilde\phi_c\} &= \{\pi_a+2\epsilon_{ab}K^b,\tilde\Pi_c\} = 2\epsilon_{ac}\\
    \{\tilde\phi_{ab},\phi_B\} &= \{\tilde\Pi_{ab}+\epsilon_{ab}B,\pi_B\} = \epsilon_{ab}
\end{eqnarray}
The derivatives of the momenta of $N,N^a,\omega_0^{0a},\omega_0^{ab}$ are:
\begin{align}
    \label{first_class_constrain}
    \dot \phi_N &= \{\pi_N,{H}_{tot}\}=\epsilon_{ab}\Big[\tilde R_2^{ab}-K^a\wedge K^b\Big]\approx0\\
    \dot \phi_{N,a}&= \{\pi_{N,a},{H}_{tot}\}=-2\e_{ab}\nabla_2K^b\approx0\\
    \dot \phi^0_{0a} &=\{\Pi^0_{0a},{H}_{tot}\}=-2\e_{ab}\nabla_2\tilde e^b\approx0\\
     \dot \phi^0_{ab} &=\{\Pi^0_{ab},{H}_{tot}\}=2\e_{c[a}\tilde\e^c\wedge K_{b]}\approx0
\end{align}
Since no Lagrange multipliers appear explicitly, these relations are natural candidates for first-class constraints.Their stability under time evolution must nevertheless be verified.

As a representative example, consider the constraint \eqref{first_class_constrain}, $\phi^0_{ab}$.
Using \eqref{time_derivative_equation}, the equations of motion for the canonical variables read 
\begin{align}
    \dot{\tilde e}^a &= \{\tilde e^a, {H}_{Tot}\} = u^a\\
    \dot K^a &= \{K^a, {H}_{Tot}\} = {\tilde U}^a\\
    \dot {\tilde \omega}^{ab} &=  \{\tilde \omega^{ab}, {H}_{Tot}\} = {\tilde U}^{ab}
\end{align}
$\ddot \phi^0_{ab}$ then become
\begin{equation}
\begin{split}
    \ddot \phi^0_{ab} =& -2\e_{ab}\{d_2\tilde e^b + \tilde\omega^b_{\;c}\wedge\tilde e^c,{H}_{Tot}\} \\
    =&-2\e_{ab}(d_2u^b + \tilde U^b_{\;c}\wedge\tilde e^c+\tilde\omega^b_{\;c}\wedge u^c) \\
    =&-2\e_{ab}(d_2\dot{\tilde e}^b + \dot{\tilde\omega}^b_{\;c}\wedge\tilde e^c+\tilde\omega^b_{\;c}\wedge\dot{\tilde e}^c)\\
    =&\partial_0(-2\e_{ab}\nabla_2\tilde e^b)
\end{split}
\end{equation}
which is satisfied if $\dot \phi^0_{ab}\approx0$.

The following two equations determine two Lagrange multipliers $u_i$
\begin{align}
     \dot \phi_a &=
    \{\pi_a,{H}_{tot}\}=-2\e_{ab}[K_c\omega_0^{cb}-\nabla_2\omega_0^{0b}]+2\e_{ab}\tilde U^b\approx0\\
    \dot {\tilde{\phi}}_{a} &=\{\tilde\Pi_a,{H}_{tot}\}=-2\e_{ab}(NK^b-\nabla_2N^b)+2\e_{cd}\omega_{0a}^{\;\;\;d}\tilde e^c-2\e_{ab}u^b\approx0\\
\end{align}
These constraints are second-class and let us recover the equations of motion,
\begin{align}
    \dot {\tilde e}^a &=  u^a = -(NK^a-\nabla_2N^a)+\e^{ba}\e_{cd}\omega_{0b}^{\;\;\;d}\tilde e^c\\
    \dot {K}^a &=  \tilde U^a = K_b\omega_0^{ba}-\nabla_2\omega_0^{0a}\\
\end{align}
The remaining consistency conditions are 
\begin{align}    
    \dot {\tilde{\phi}}_{ab} &=\{\tilde\Pi_{ab},{H}_{tot}\}=-\e_{ab}(\nabla_2N+\dot B)-2\e_{[a|d}N_{|b]}K^d+\e_{c[a|}\omega_{0\;\;|b]}^{\;0}\tilde e^c+\e_{ab}u^B\approx0\\
    \dot \phi_B &= \{\pi_B,{H}_{tot}\}=\e_{ab}[\partial_0(\tilde\omega^{ab})-\tilde U^{ab}]\approx0\\
    \dot \phi_c &= \{\pi_c,{H}_{tot}\}=-\partial_0\nabla_2{\overline{c}}\approx0\\
    \dot {\phi_{\overline{c}}} &= \{\pi_{\overline{c}},{H}_{tot}\}=-\partial_0\nabla_2(c) \approx0
\end{align}
$ \dot {\tilde{\phi}}_{ab}\approx 0$ is a second-class constraint since it defines the Lagrange multiplier $u_B$; after contracting with $\e^{ab}$ we get
\begin{equation}
    \nabla_2N+\dot B+N_bK^d-\omega_0^{0b}\tilde e_b = u^B
\end{equation}
Although, since $u^B=\dot B$, this constraint can be written as 
\begin{equation}
    \nabla_2N+N_bK^d-\omega_0^{0b}\tilde e_b = 0
\end{equation}
which is the component of $\nabla e^0$ orthogonal to the Cauchy surface, and it is coherent with the covariant description above. 

The equation $\dot \phi_B$ does not generate additional information, since it is already known that $\partial_0(\tilde\omega^{ab})=\tilde U^{ab}$.
For the same reason, the equations for the ghost and the antighost do not give us any new information.
The (anti)ghost equations show that the gauge fixing reduces the local Lorentz symmetry from SO(2,1) to its abelian subgroup SO(2). In this reduced setting, the commutator between any two elements of the algebra vanishes, since SO(2) is one-dimensional. Consequently, the covariant double derivative of the ghost field satisfies $\nabla_2 c = [\tilde R_2,c]=0$ 

Finally, after some manipulation and solving the second class constraint, the first class constraints are
\begin{align}
    \label{constrain}
    &\epsilon_{ab}\Big[\tilde R_2^{ab}-K^a\wedge K^b\Big] = \epsilon_{ab}\Big[\tilde R_2^{ab}-\frac{1}{4}\pi^a\wedge \pi^b\Big]\approx0\\
    &\nabla_2K^a= \nabla_2\pi^a\approx0\\
    &\nabla_2\tilde e^a\approx0\\
    &\tilde\e^a\wedge K_{a}= \tilde\e^a\wedge \pi_{a}\approx0
\end{align}
For each constrain, a 0-form Lagrangian multiplier $\{A_N,A_{N,a},A_{\omega,a},A_\omega\}$ and its corresponding 0-form ghosts  $\{c_N,c_{N,a},c_{\omega,a},c_\omega\}$ are introduced. In addition, to implement the gauge fixing condition and the corresponding ghost dynamics, BRST doublets 2-form $\{\rho^N,\rho^{N,a},\rho^{\omega,a},\rho^\omega\}$  and its corresponding 2-form ghosts  $\{\overline{c}^N,\overline{c}^{N,a},\overline{c}^{\omega,a},\overline{c}^\omega\}$ is needed. After labeling with the generic indices $i,j$ the constraint, those degrees of freedom are quantized as follows 
\begin{align}
    &[A_i(x),\rho^j(y)] \;\;= i \delta^{j}_i\delta^2(x-y)Vol_\Sigma\nonumber\\
    &[c_i(x),p^j(y)]_+ \;=  \delta^{j}_j\delta^2(x-y)Vol_\Sigma\nonumber\\
    &[\overline{c}^i(x),\overline p_{j}(y)]_+ =  \delta^{i}_j\delta^2(x-y)Vol_\Sigma\nonumber
\end{align}
The constraints \eqref{constrain} form a closed algebra under the Poisson bracket, given by
\begin{align}
    &\{\epsilon_{ab}\Big[\tilde R_2^{ab}-\frac{1}{4}\pi^a\wedge \pi^b\Big],\nabla_2\pi^c\}=0 \quad\;\; \{\epsilon_{ab}\Big[\tilde R_2^{ab}-\frac{1}{4}\pi^a\wedge \pi^b\Big],\nabla_2\tilde e^c\} =0\\ 
    &\{\epsilon_{ab}\Big[\tilde R_2^{ab}-\frac{1}{4}\pi^a\wedge \pi^b\Big],\tilde\e^c\wedge \pi_{c}\}=0 \quad\{\nabla_2\tilde e^a,\tilde\e^b\wedge \pi_{b}\} = \nabla_2\tilde e^a\delta^2(x-y)\nonumber\\
    &\{\nabla_2\pi^a,\nabla_2\tilde e^b\} = 0 \quad\quad\quad\quad\quad\quad\quad\;\quad \{\nabla_2\pi^a,\e^b\wedge \pi_{b}\} = \nabla_2\pi^a\delta^2(x-y)\nonumber    
\end{align}
The BRST charge $Q$ is then
\begin{equation}
\begin{split}
        Q = &\int_\Sigma \Big\{c_N\epsilon_{ab}\Big[\tilde R_2^{ab}-\frac{1}{4}\pi^a\wedge \pi^b\Big] +c_{N,a} \nabla_2\pi^a + c_{\omega,a}\nabla_2\tilde e^a+c_\omega\e^a\wedge \pi_{a} 
        + c_{N,a}c_\omega p^{N,a}+c_{\omega,a}c_\omega p^{\omega,a}  \\&+\overline{p}_N\rho^N+\overline{p}_{N,a}(\rho^{N,a}+c_\omega\overline{c}^{N^a})+\overline{p}_{\omega,a}(\rho^{\omega,a}+c_\omega\overline{c}^{\omega,a} ) +\overline{p}_\omega\rho^\omega+    c_{N,a}A_\omega\rho^{N,a}+c_{\omega,a}A_\omega\rho^{\omega,a}\Big\}
\end{split}
\end{equation}
which is nilpotent due to the first-class algebra of constraints and the canonical commutation relations. This expression can be ulteriorly simplified by a redefinition of $\rho^i \rightarrow \rho^i+f_{\;\;k}^{ij}c_j\,\overline{c}^k$ to get 

\begin{equation}
\begin{split}
        Q = &\int_\Sigma \Big\{c_N\epsilon_{ab}\Big[\tilde R_2^{ab}-\frac{1}{4}\pi^a\wedge \pi^b\Big] +c_{N,a} \nabla_2\pi^a + c_{\omega,a}\nabla_2\tilde e^a+c_\omega\e^a\wedge \pi_{a} 
        + c_{N,a}c_\omega p^{N,a}+c_{\omega,a}c_\omega p^{\omega,a}  \\&+\overline{p}_N\rho^N+\overline{p}_{N,a}\rho^{N,a}+\overline{p}_{\omega,a}\rho^{\omega,a} +\overline{p}_\omega\rho^\omega\Big\}
\end{split}
\end{equation}
This allows us to decompose $Q$ into its linear and non-linear parts, namely
\begin{align}
    Q_0 &= \int_\Sigma \Big\{c_N\epsilon_{ab}d_2\tilde\omega^{ab} +c_{N,a} d_2\pi^a + c_{\omega,a}d_2\tilde e^a +     \overline{p}_N\rho^N+\overline{p}_{N,a}\rho^{N,a}+\overline{p}_{\omega,a}\rho^{\omega,a} +\overline{p}_\omega  \rho^\omega \Big\}  \\
    Q_1&=\int_\Sigma \Big\{c_N\e_{ab}\Big[\tilde \omega^{ac}\wedge\tilde\omega_c^{\;b}-\frac{1}{4}\pi^a\wedge \pi^b\Big] +c_{N,a} \tilde\omega^a_{\;b}\wedge\pi^b + c_{\omega,a}\tilde\omega^a_{\;b}\wedge\tilde e^b+c_\omega\e^a\wedge \pi_{a}
    \\& \quad \quad c_{N,a}c_\omega p^{N,a}+c_{\omega,a}c_\omega p^{\omega,a}\Big\}
    \nonumber
\end{align}
The following fields are defined so that they are conjugate to the linear part of the constraints:
\begin{align}
    \Phi_N(y) &=\frac{1}{2}\int_z  G(y-z) \e^{ab}d_{z}\star\tilde\Pi_{ab}\\
    \Phi_{N,a}(y) &=\int_z  G(y-z) d_{z}\star\tilde \e^a\\
    \Phi_{\omega,a}(y) &=\int_z G(y-z)d_{z}\star\tilde \pi^a
\end{align}
These fields satisfy the straightforward BRST relations:
\begin{equation}
    \big[Q_0,\Phi_N(y)\big]= i c_N(y) \quad \big[Q_0,\Phi_{N,a}(y)\big]=- ic_{N,a}(y) \quad  \big[Q_0,\Phi_{\omega,a}(y)\big]= ic_{\omega,a}(y)
\end{equation}
An operator $R$  can then be defined such that its anticommutator with $Q_0$ gives
\begin{equation}
        [Q_0,R]_+ = iS   
\end{equation}
with $S$ the number counting operator that assigns to each field the corresponding ghost number, for instance, assigning +1 to each ghost and -1 to each anti-ghost. 
It is easy to check that $R$ is 
\begin{equation}
    R = \int_\Sigma\Big( -\overline{c}^N A_N - \overline{c}^{N,a}
    A_{N,a}-\overline{c}^{\omega,a}  A^{\omega,a}+p^N\Phi_N  +p^{N,a}\Phi_{N,a}  -p^{\omega,a} \Phi_{\omega,a} \Big) 
\end{equation}
and 
\begin{align}
    S = \int_\Sigma& \Big(\overline{p}_N\overline{c}^N + \overline{p}_{N,a}\overline{c}^{N,a} + c_Np^N + c_{N,a}p^{N,a}+i\rho^NA_N+i\rho^{N,a}A_{N,a}
    \nonumber \\
    & -i(\epsilon_{ab}d_2\tilde\omega^{ab})\Phi_N  -i( d_2\pi^a)\Phi_{N,a} 
    +\overline{p}_{\omega,a}\overline{c}^{\omega,a} +  c_{\omega,a}p^{\omega,a} +i\rho^{\omega,a}A_{\omega,a} +i( d_2\tilde e^a)\Phi_{\omega,a}\Big) 
\end{align}
With these building blocks, the intertwiner $\Omega$ can be computed by solving the differential equation
\begin{equation}
    \frac{\partial\Omega}{\partial t} = i\Omega(t)[Q_1,R]_+
\end{equation}
with the commutator given, in this case, by 
\begin{align}
     \Gamma \equiv  [Q_1,R]_+ =& \int_\Sigma\e_{ab}\Big(\tilde\omega^a_{\;c}\wedge\tilde\omega^{cb}-\frac{1}{4}\pi^a\wedge\pi^b\Big)\Phi_N+\tilde\omega^a_{\;b}\wedge\pi^b\Phi_{N,a}-\tilde\omega^a_{\;b}\wedge e^b\Phi_{\omega,a}\nonumber\\
     &-\frac{i}{2}\int_{\Sigma}\int_{\Sigma'}d_x\star_x \e^{ab}\Big(c_{\omega,a}(x)\tilde e_b(x) + c_{N,a}(x)\pi_b(x)\Big)G(y-x)p^N(y)\\
     &-i\int_{\Sigma}\int_{\Sigma'}d_x\star_x\Big(\frac{\e_{ab}}{2}c^N(x)\pi^a(x)+c_{N,a}(x)\tilde\omega^a_{\;b}(x)+c_\omega(x)\tilde e_b\Big)G(y-x)p^{N,b}(y)\nonumber\\
     &+i\int_{\Sigma}\int_{\Sigma'}d_x\star_x\Big(c_{\omega,a}(x)\tilde\omega^a_{\;b}(x)+c_\omega(x)\pi_b\Big)G(y-x)p^{\omega,b}(y)\nonumber
\end{align}

The intertwiner is therefore $\Omega(t) = e^{i\Gamma t}$.

\subsection{2+1 Chern-Simons Gravity with Cosmological Constant}

In 3D AdS gravity with negative cosmological constant $\Lambda = -1/{l}^2$ the Gauge group $SO(2,2)$ can be splitted into two indipendent sector since $SO(2,2) \cong SL(2,\mathbb{R})\times SL(2,\mathbb{R})$, therefore the triad and the spin connection  combine into two $SL(2,\mathbb{R})$ gauge fields: 
\begin{equation}
    \label{eq!}
    A^a_L = \omega^a +\frac{e^a}{l} \qquad A^a_R = \omega^a -\frac{e^a}{l}
\end{equation}
with $\omega^a = \frac12 \e^a_{\;bc}\omega^{bc}$

With a positive cosmological constant, this splitting can still be done since the gauge group is $SO(3,1) \cong SL(2,\mathbb{C})$, but now
\begin{equation}
    A^a = \omega^a +\frac{e^a}{l} \qquad \overline A^a = \omega^a -\frac{e^a}{l}
\end{equation}
are complex conjugates, so they are no longer independent.
In both cases, the action is separated into independent components
\begin{align}
    &S_{eh} = \frac{1}{4\sqrt{-\Lambda}}\int_{\mathcal{M}_3}\bigg(A^a_L\wedge dA_{aL} +\frac{2}{3}\e_{abc}A_L^a\wedge A_L^b\wedge A_L^c\bigg)-\bigg(A^a_R\wedge dA_{aR} +\frac{2}{3}\e_{abc}A_R^a\wedge A_R^b\wedge A_R^c\bigg)\\
    &S_{eh} = \frac{i}{4\sqrt\Lambda}\int_{\mathcal{M}_3}\bigg(A^a\wedge dA_{a} +\frac{2}{3}\e_{abc}A^a\wedge A^b\wedge A^c\bigg)-\bigg(\overline A^a\wedge d\overline A_{a} +\frac{2}{3}\e_{abc}\overline A^a\wedge \overline A^b\wedge \overline A^c\bigg)
\end{align}

Firstly, we are going to focus on the construction of the BRST charge for $AdS_3$ gravity, and we see that it's a pretty straightforward process; Since the action decomposes into two independent Chern-Simons theories, if the assumption that $\Sigma$ allows a simplectic structure holds, it is possible to build for each independent sector the following conjugate variables 
\begin{equation}
    [A_z^{aL} (z_1), A^{bL}_{\overline z} (z_2)] = i\sqrt{\Lambda}g^{ab}\delta^2(z_1 - z_2) \qquad [A_z^{aR} (z_1), A^{bR}_{\overline z} (z_2)] = -i\sqrt{\Lambda}g^{ab}\delta^2(z_1 - z_2)
\end{equation}
In particular, the holomorphic polarization for the Left sector and the antiholomorphic polarization for the Right sector are chosen, so that the wavefunction is in the form $\Psi(A^L_z, A^R_{\overline{z}})$.

The Hamiltonian density is given by
\begin{equation}
    \mathcal{H} = \frac{1}{4\sqrt{\Lambda}}(-A_{L,0}^a\mathcal{G}^L_a + A_{R,0}^a\mathcal{G}^R_a) 
\end{equation}
where $\mathcal{G}$ is the Gauss constraint that assumes the following form for both the left and the right sector
\begin{equation}
    \mathcal{G}^a = \partial A^a_{\overline{z}}- \overline{\partial}A^a_z + \e^a_{bc}A_z^bA_{\overline{z}}^c
\end{equation}

The BRST operator is given by the difference between the Left- and right-handed BRST charges.
\begin{equation}
    Q_{BRST} = -Q^L_{BRST} + Q^R_{BRST}
\end{equation}
where the left and right ghosts satisfy the usual algebra, and the BRST charge is
\begin{equation}
    Q = \int_{\Sigma} 
    \left( c_a (\partial_z  A^{Ia}_{\bar z} - \partial_{\bar z} A^{Ia}_z + \e^a_{bc}
    A^{Ib}_z A^{Ic}_{\bar z}) + \rho^{Ia} \pi^{Ib}_a + \frac12 f^a_{bc} \pi^{Ic}_a  c^{Ib} c^{Ic}
    \right) \nonumber \\
\end{equation}
where I span between $I = (L,R)$

Following the same procedure of section [3.2], and knowing that the quantities of the right sector commute with the ones in the left sector, it is found that
\begin{equation}
    \label{intLMax}
    \Gamma = -\Gamma_L+\Gamma_R
\end{equation}
where sigma is given by \eqref{holF}.
The intertwiner, therefore, can then be written as 
\begin{equation}
    \Omega = e^{i(-\Gamma_L+\Gamma_R)}
\end{equation}

In the case of a positive cosmological constant, we no longer have two separated sectors since the reality condition imposes that 
\begin{equation}
    \overline{A} = A^\dagger.
\end{equation}
Nevertheless, one way to deal with this system is to treat $A$ and $\overline{A}$ as two separate connections and impose the reality condition at the very end.
If we treat $A$ and $\overline{A}$ as two independent systems, inspired by the negative cosmological constant case, we will have that the Hamiltonian density can be written as 
\begin{equation}
    \mathcal{H} = \frac{1}{4\sqrt{\Lambda}}(-A_{0}^a\mathcal{G}_a + \overline A_{0}^a\overline {\mathcal{G}}_a) 
\end{equation}
where $\mathcal{G}$ and $\overline {\mathcal{G}}$ are the Gauss constrains for the connection $A$ and $\overline{A}$ respectively. 

In this case, the conjugate variables are 
\begin{equation}
    [A^a_z(z_1), A^b_z(z_2)\}] = \frac{2\pi i}{k}\delta^{ab}\delta^2(z_1 - z_2) \qquad  [\overline A^a_z(z_1), \overline A^b_z(z_2)\}] = -\frac{2\pi i}{k}\delta^{ab}\delta^2(z_1 - z_2)
\end{equation}
When we have set for simplicity $k = \frac{i}{4\sqrt{\Lambda}}$.  Then we can solve the reality condition; the commutators are already consistent, and we get that $\overline{\mathcal{G}} = \mathcal{G}^\dagger$ 

To build the BRST charge, a complex pair of (anti) ghosts and their momenta is needed.

The BRST charge becomes
\begin{equation}
    Q_{BRST} = -\int_{\Sigma} 
    \left( c_a (\partial_z  A^{a}_{\bar z} - \partial_{\bar z} A^{a}_z + \e^a_{bc}
    A^{b}_z A^{c}_{\bar z}) + \rho^{a} \pi^{b}_a + \frac12 f^a_{bc} \pi^{c}_a  c^{b} c^{c} 
    \right) +hc \nonumber \\
\end{equation}

We can now compute the intertwiner 
\begin{equation}
    \Omega(t) = e^{i(-\Gamma + \Gamma^\dagger)t}
\end{equation}
where $\Gamma$ is given by \eqref{holF}.

We now aim to compare these results with those derived in the absence of the mapping from general relativity to Chern–Simons theory. 
The action is given by
\begin{align}
    S = &\int_{\mathcal{M}_3} \e_{abc}\big(e^a\wedge R^{bc}(\omega)-\frac{\Lambda}{3}e^a\wedge e^b\wedge e^c\big)\\
    &\int dt \int_\Sigma  \e_{abc}\big[e^a_0 (R^{bc}_2(\hat{\omega})-\Lambda \hat{e}^b\wedge\hat{e}^c) -\omega^{bc}_0\nabla_2 \hat e^a -\hat{e}^a\wedge \partial_0 \hat\omega^{bc}\big]\nonumber
\end{align}
Following a procedure analogous to that adopted in general relativity in the absence of a cosmological constant, one obtains the following algebra of first-class constraints:
\begin{equation}
    \{\dot\pi^0_a,\dot\pi^0_b\} = 4\Lambda\e_{abc}\nabla_2\hat e^c \qquad \{\nabla_2 \hat e^{a}, \nabla_2 \hat e^{b} \} = -\e^{abc}\nabla_2 \hat e_c \qquad \{\dot\pi^{0,a},\nabla_2\hat e^{b}\} = \e^{abc}\dot\pi^0_c
\end{equation}
With the first-class constraint defined as follows:
\begin{equation}
    \dot\pi^0_a = \e_{abc}(R_2^{bc}(\hat{\omega})-\Lambda \hat{e}^b\wedge\hat{e}^c) \qquad
\end{equation}.
The presence of the cosmological constant introduces a term such that the curvature no longer commutes with itself; rather, a contribution proportional to the torsion appears.
The BRST charge is modified as follows
\begin{align}
    Q = \int_\Sigma &( \epsilon_{abc} c^a (R^{bc}(\hat \omega)-\Lambda\hat e^b\wedge\hat e^c) + 
    \epsilon_{abc} \lambda^{ab} T^c(\hat e, \hat \omega) 
    + \rho_a \pi^{b,a} + \rho_{ab} \pi^{b,ab} + \nonumber\\
    &\frac12\pi^{c,}_{ab} \lambda^{ac}  \lambda_{c}^{~b} +
     \pi^{c,}_{a}\lambda^{ab}c_b  -\Lambda\e_{ab}^{\;\;c}\pi^{c,}_{c}c^ac^b
\end{align}
We can split the BRST charge in the following way:
\begin{eqnarray}
Q_0 &=& \int_\Sigma \left( \epsilon_{abc} c^a d \hat \omega^{bc} + 
    \epsilon_{abc} \lambda^{ab} d \hat e^a 
       + \rho_a \pi^{b,a} + \rho_{ab} \pi^{b,ab}\right), \nonumber \\
Q_1 &=& \int_\Sigma \left( \epsilon_{abc} c^a (\hat\omega^{bd} \wedge \hat\omega_{d}^{~~c})+ 
    \epsilon_{abc} \lambda^{ab} \hat \omega^{cd} \wedge \hat e_d + \frac12\pi^{c,}_{ab} \lambda^{ac}  \lambda_{c}^{~b} +
    \pi^{c,}_{a}\lambda^{ab}c_b    \right) \\
    Q_{-3} &=& -\Lambda\int_\Sigma \e_{abc}(c^a\hat e^b\wedge\hat e^c +  \pi^{c,}_{ab}c^{a}c^{b})   \nonumber
\end{eqnarray}
Again, in the present case, we run in the situation where the BRST charge has a negative-charged piece and the two-step procedure has to be adopted to built the intertwiner as we are illustrate below.

\subsection{Intertwiner}

\newcommand{\QL}{Q_{\Lambda=0}}

The BRST charge has the following form 
\begin{eqnarray}
    \label{IIA}
    Q_\Lambda = Q_0 + Q_1 + Q_{-3} = Q_{\Lambda=0} + Q_{-3} 
\end{eqnarray}
where $\QL =  Q_0 + Q_1$ is nilpotent BRST charge for vanishing 
cosmological constant. The different pieces satisfy the relations 
$\{\QL, Q_{-3}\} =0$ and $\{Q_{-3}, Q_{-3}\} =0$. In addition, due to the different charges we have also the refined commutation relations $\{Q_0, Q_{-3}\} = \{Q_1, Q_{-3}\} =0$. The former implies that 
\begin{eqnarray}
    \label{IIB}
    Q_{-3} &=& [Q_0, \Gamma_\Lambda]\,, \nonumber \\
    \Gamma_\Lambda &=& \int_\Sigma \left(\epsilon_{abc} \Phi^a \hat e^b \wedge \hat e^c - \pi^{c,}_{ab} \Phi^a c^b
    +  \epsilon_{abc} \Phi^a d \Phi^b \wedge \hat e^c 
    - \epsilon_{abc} \Phi^a d \Phi^b \wedge d\Phi^c   
    \right)
\end{eqnarray}

Using $\{Q_1, Q_{-3}\} =0$, one has 
$0= \{Q_1, [Q_0 \Gamma_\Lambda]\} = \{Q_0, [Q_1 \Gamma_\Lambda]\}$. Since the cohomology of $Q_0$ vanishes, it holds that 
\begin{eqnarray}
    \label{IIC}
    [Q_1, \Gamma_\Lambda] =     [Q_0, \Gamma^{(1)}_\Lambda]
\end{eqnarray}
We need to find an expression $P_\Lambda$ such that $Q_{-3} = [\QL, P_\Lambda]$ in order to get the intertwiner where $Q_0$ is replaced by $\QL$. 
For that, one adds and subtracts two pieces to the above expression 
\begin{eqnarray}
    \label{IID}
   Q_{-3} = [Q_0 + Q_1, \Gamma_\Lambda] -   [Q_0, \Gamma^{(1)}_\Lambda]
\end{eqnarray}
Acting with $Q_1$ on \eqref{IIC} yield
\begin{eqnarray}
    \label{IIE}
    [Q_1, \Gamma^{(1)}_\Lambda] =     [Q_0, \Gamma^{(2)}_\Lambda]
\end{eqnarray}
Repeating the computation in the same way, one obtains 
\begin{eqnarray}
    \label{IIF}
    Q_{-3} = \left[Q_0 + Q_1, \sum_{l=0}^\infty (-1)^l \Gamma^{(l)}_\Lambda\right] 
\end{eqnarray}
with $\Gamma^{(0)}_\Lambda = \Gamma_\Lambda$. 
If is set $P_\Lambda  = \sum_{l=0}^\infty (-1)^l \Gamma^{(l)}_\Lambda$, one has $Q_{-3} = [\QL, P_\Lambda]$. 
Then, it holds 
\begin{eqnarray}
    \label{IIG}
   Q_\Lambda = \QL + \Big[\QL, P_\Lambda\Big] 
\end{eqnarray}
Now, one would like to exponentiate the above expression 
\begin{eqnarray}
    \label{IIH}
   Q_\Lambda = e^{-P_\Lambda}\QL e^{P_\Lambda} 
\end{eqnarray}
and for that, the vanishing of 
$\Big[\Big[\QL, P_\Lambda \Big],P_\Lambda \Big]$ is needed. 
This can be checked as follows: since $\{Q_{-3}, Q_{-3}\} =0$,  
\begin{eqnarray}
    \label{IIL}
    \Big[P_\Lambda, \Big[P_\Lambda, \QL \Big], \QL \Big] =0
\end{eqnarray}
Following the construction outlined in Section 2, the exponent $P_\Lambda$ can be redefined to reabsorb the additional terms coming from the exponentiation. 
Finally, using the construction of the intertwiner $\Omega$ for $\QL$, it is possible to write the full BRST charge as 
\begin{eqnarray}
\label{IIM}
Q_\Lambda = e^{-P_\Lambda} e^{-\Gamma} Q_0  e^{\Gamma} e^{P_\Lambda}
\end{eqnarray}
and since $P_\Lambda$ and $\Gamma$ have non-trivial commutation relations, it is not convenient to bring them to an exponential of a single operator using the BHC formula. This shows the big difference between the holomorphic quantization technique versus the canonical quantization. The full intertwiner is definitely more complicated in the latter case. Note that by setting $\Lambda =0$, the formula already obtained in the previous sections is retrieved. The present proof is based on the vanishing of BRST cohomology on local fields of $Q_0$. $\Phi^a$ and $\Phi^{ab}$ are non-local, but it can be shown that extending the analysis of the cohomology to those fields, the cohomology can be studied along the lines discussed in the case of Chern-Simons theory. 

. 




\section{Outlook}

In this paper, we construct the intertwiner $\Omega$
for several $D=3$ models in which quantization is feasible, and BRST symmetry plays a crucial role. We analyze these models from different perspectives in order to highlight the difficulties encountered in the construction. The present work serves a twofold purpose: first, to provide general tools for building a framework applicable to higher-dimensional models (such as gravity and supergravity in 
$D=4$); and second, to study observables both for the models considered here and for their higher-dimensional counterparts. Both directions will be pursued in future work.

\section*{Acknowledgments} 
We thank M. Porrati for several important discussions, helping us with insights and ideas on the subject. 
We would like to thank S. Cacciatori for discussions on D=3 gravity and G. Barnich and E. de Sabbata for interesting discussion on the cohomology. P.A.G. would like to thank the CERN Th-Department, where this work has been completed, and he is partially supported by HORIZON-MSCA-2021-SE-01-101086123 CaLIGOLA.

\section*{Appendix A: Local Cohomologies}

Let us call \emph{local forms} the differential forms $\Omega_p^q$ whose coefficients
$\omega_{\mu_1\ldots\mu_p}(x)$ are local functionals, i.e.\ polynomials in the fields
and their derivatives, all taken at the same spacetime point $x$. 
The index $q$ denotes the ghost number. 
The integral of an $x$-dependent local form is also called a \emph{local functional}, for that, we have functionals 
\begin{eqnarray}
    \label{funcA}
    \Omega[\varphi] = \int_{\mathcal{M}^p} 
    \omega_{\mu_1\ldots\mu_p}[\varphi(x)] dx^{\mu_1} \wedge \dots \wedge dx^{\mu_p}\,. 
\end{eqnarray}
Closed functionals $\Big[Q,  \Omega[\varphi]\Big]=0$ corresponds to 
closed local functional up to a total derivative 
\begin{eqnarray}
    \label{funcB}
    \Big[Q, \omega^q_{\mu_1\ldots\mu_p}[\varphi(x)]\Big] = 
    \partial_{[\mu_1}  \omega^{q+1}_{\mu_2\ldots\mu_p]}[\varphi(x)]
\end{eqnarray}
and in the same way, exact functional $ \Omega[\varphi] = \Big[Q, \Xi[\varphi]\Big]$ corresponds to exact/modulo total-derivative local-functionals $\omega^q_{\mu_1\ldots\mu_p}[\varphi(x)] = [Q, \omega^{q-1}_{\mu_1\ldots\mu_p}[\varphi(x)]\Big] +  \partial_{[\mu_1}  \omega^{q}_{\mu_2\ldots\mu_p]}[\varphi(x)]$. In the form language we use the notation $\Omega^q_p$ and $H(Q|d)$ to denote the 
\emph{local cohomologies}. The problem of solving the (local) cohomology of $Q$ (denoted by $H(Q)$ or that of $Q$ modulo $d$ (denoted by $H(Q|d)$ may be rather difficult in general.  The following proposition allows one to replace the coboundary operator $Q$ by a simpler one, $Q_0$ (see \cite{Dixon:1991wi,Brandt:1989rd,Brandt:1989gy,Piguet:1995er}). 

Let $S$ be a filtration operator mapping the space of forms $\Omega^q_p$ into itself, the eigenvalues of $S$ being nonnegative integers. Suppose that the forms $\Omega^q_p$ and the coboundary operator $Q$ admit expansions according to these eigenvalues:
\begin{equation}
\label{funcBB}
\Omega^q_p = \sum_{n\geq 0} \Omega^q_{p, n},~~~~~
Q= \sum_{n\geq 0} Q_n, ~~~~~
[S,Q_n]=n\,Q_n\,, ~~~~~
[S, \Omega^q_{p, n}]=n \,  \Omega^q_{p, (n)}\,. 
\end{equation}
Assuming that the filtration operator $S$ commutes with the exterior
derivative $d$. Then:
\begin{enumerate}
\item $Q_0$ is a coboundary operator, i.e.\ $Q_0^2=0$.
\item The cohomology of $Q$ is isomorphic to a subspace of the cohomology of $Q_0$, namely $H(Q) \sim H'(Q_0) \subset H(Q_0)$.  
\item The cohomology of $Q$ modulo $d$ is isomorphic to a subspace of the cohomology of $Q_0$ modulo $d$, namely $H(Q|d) \sim H'(Q_0|d) \subset H(Q_0)$. 
\end{enumerate}

The selection of the most suitable filtration operator $S$
depends on the specific problem being studied. Following \cite{Brandt:1989rd,Brandt:1989gy,Piguet:1995er}, 
the proof of point (1) is immediate. We prove point (3); the proof of (2) follows.

Let $\Omega^q_p$ be a representative of $H(Q|d)$. 
One can always choose $\Omega^q$ such that
$\Omega^q_p = \sum_{n\geq N} \Omega^q_{p,n}$, 
with a leading term $\Omega^q_{p,N}$ which is nonvanishing and belongs to the $H(Q_0|d)$. Indeed, the cocycle condition
$[Q,\Omega^q_p] + d \Omega^{q+1}_{p-1} = 0$ implies, at lowest order, 
$[Q_0,\Omega^q_{p,N}] + d \Omega^{q+1}_{p-1,N} = 0$. 
If $\Omega^q_{p, N}$ is trivial in the $Q_0$-cohomology modulo $d$, i.e. $\Omega^q_{p,N} = [Q_0, \Theta^{q-1}_{p, N}] + d \Theta^{q}_{p-1, N},$
one can redefine $\Omega^q_{p, N}$ by subtracting $
[Q, \Theta^{q}_{p, N}] + d \Theta^{q-1}_{N} + {\rm terms~ of~ order~ n \geq N + 1}$,
thereby raising the lowest filtration order. Repeating this procedure yields a representative whose leading term is nontrivial in the $Q_0$-cohomology modulo $d$.

We must still show that the map from the cohomology of $Q$ modulo $d$ to the cohomology of $Q_0$ modulo $d$ thus introduced can be defined
as an \emph{injective} map, i.e.\ a map such that if $\Omega^q_p$ and
$\widetilde{\Omega}^q_p$ are representatives of two distinct cohomology classes of $Q$ modulo $d$, then their leading terms
$\Omega^q_{p, N}$ and $\widetilde{\Omega}^q_{p, N'}$ belong to two distinct cohomology classes of $Q_0$ modulo $d$. 
This is obvious if $N \neq N'$. For $N=N'$, suppose the contrary, i.e.\
\begin{equation}
\widetilde{\Omega}^q_{p,N} - \Omega^q_{p,N}
= [Q_0, \Xi^{q-1}_{p,N}] + d \Xi^{q}_{p-1,N}\,.
\end{equation}

Then, 
\begin{equation}
\widetilde{\Omega}^q_p - \Omega^q_p
=
Q \Xi^{q-1}_p + d \Xi^{q}_{p-1}
+
\sum_{n \ge N+1} \Psi^q_{p,n} .
\end{equation}

This allows one to choose, for example, $\Omega^q_p$ and
$\widehat{\Omega}^q_p
=
\widetilde{\Omega}^q_p
-
[Q, \Xi^{q-1}_p]
-
d \Xi^{q}_{p-1}$
as independent representatives.
Repeating the operation if necessary, it is then clear that one arrives
at a pair of representatives whose leading terms are not equivalent in
the $Q_0$-cohomology modulo $d$, since the lowest filtration degree
of $\widehat{\Omega}^q$ is strictly greater than $N$.
This completes the proof.





\end{document}